\documentclass[
    aps,
    prd,
    reprint,
    superscriptaddress,
    amsmath,
    amssymb,
    floatfix
    ]{revtex4-2}

\usepackage[utf8]{inputenc}
\usepackage[T1]{fontenc}

\usepackage{graphicx}       
\usepackage{dcolumn}        
\usepackage{bm}             

\usepackage[
    colorlinks=true,
    linkcolor=blue,
    citecolor=blue,
    urlcolor=blue
    ]{hyperref}
    
\usepackage{booktabs}
\usepackage{multirow}
\usepackage{amsmath, amssymb}
\usepackage{graphicx}
\usepackage{braket}
\usepackage{geometry}
\usepackage[english]{babel}
\usepackage{hyperref}
\usepackage{color}
\usepackage{wasysym}

\def\rms{{\rm s}}

\def\sl{\rms_\lambda}

\def\proof{\noindent{\sl Proof:}\kern0.6em}

\def\frac#1#2{\hbox{$#1\over#2$}}
\def\dual{\mathstrut^*\kern-0.1em}

\def\lvec#1{\setbox0=\hbox{$#1$}
    \setbox1=\hbox{$\scriptstyle\leftarrow$}
    #1\kern-\wd0\smash{
    \raise\ht0\hbox{$\raise1pt\hbox{$\scriptstyle\leftarrow$}$}}
    \kern-\wd1\kern\wd0}
\def\rvec#1{\setbox0=\hbox{$#1$}
    \setbox1=\hbox{$\scriptstyle\rightarrow$}
    #1\kern-\wd0\smash{
    \raise\ht0\hbox{$\raise1pt\hbox{$\scriptstyle\rightarrow$}$}}
    \kern-\wd1\kern\wd0}
\def\slash#1{\setbox0=\hbox{$#1$}\setbox1=\hbox{$\kern1pt/$}
    #1\kern-\wd0\kern1pt/\kern-\wd1\kern\wd0}

\def\nabstar#1{{\nabla\kern0.5pt\smash{\raise 4.5pt\hbox{$\ast$}}
               \kern-5.5pt_{#1}}}

\def\nabbarstar#1{{\overleftarrow{\nabla}\kern0.5pt\smash{\raise 4.5pt\hbox{$\ast$}}
               \kern-5.5pt_{#1}}}

\def\nabdbarstar#1{{\overleftrightarrow{\nabla}\kern0.5pt\smash{\raise 4.5pt\hbox{$\ast$}}
               \kern-5.5pt_{#1}}}

\def\drvstar#1{{\partial\kern0.5pt\smash{\raise 4.5pt\hbox{$\ast$}}
               \kern-6.0pt_{#1}}}

\def\ldrvstar#1{{\lvec{\,\partial}\kern-0.5pt\smash{\raise 4.5pt\hbox{$\ast$}}
               \kern-5.0pt_{#1}}}

\def\MSbar{\overline{\rm MS\kern-0.5pt}\kern0.5pt}

\def\zetabar{\bar{\zeta}}
\def\zetaprime{\zeta\kern1pt'}
\def\zetabarprime{\zetabar\kern1pt'}

\def\diracstar#1#2{
    \setbox0=\hbox{$\gamma$}\setbox1=\hbox{$\gamma_{#1}$}
    \gamma_{#1}\kern-\wd1\kern\wd0
    \smash{\raise4.5pt\hbox{$\scriptstyle#2$}}}

\def\Ds{D_{\rm s}}
\def\DsdagDs{\Ds{\Ds}^{\kern-1pt\dagger}}

\def\avg#1{{\kern1.0pt\overline{\kern-1.0pt#1\kern-1.0pt}\kern1.0pt}}

\newcommand{\be}{\begin{equation}}
\newcommand{\ee}{\end{equation}}
\newcommand{\bea}{\begin{eqnarray}}
\newcommand{\eea}{\end{eqnarray}}

\newcommand{\msbar}{{\rm \overline{MS\kern-0.05em}\kern0.05em}}

\newcommand{\ba}{\begin{eqnarray}}
\newcommand{\ea}{\end{eqnarray}}

\renewcommand{\vec}[1]{\boldsymbol{#1}}

\usepackage{xcolor}

\begin{document}
\begin{flushright}
CERN-TH-2026-230
\end{flushright}

\title{Non-perturbative renormalization of the energy-momentum tensor in the 2d $O(3)$ non-linear $\sigma$-model}

\author{Leonardo Giusti}
 \affiliation{Department of Physics "Giuseppe Occhialini", University of Milano-Bicocca, Piazza della Scienza 3, I-20126 Milano, Italy}
 \affiliation{INFN Milano-Bicocca, Piazza della Scienza 3, I-20126 Milano, Italy}
 \affiliation{Theoretical Physics Department, CERN, 1211 Geneva 23, Switzerland}

\author{Mitsuaki Hirasawa}
 \affiliation{Department of Physics "Giuseppe Occhialini", University of Milano-Bicocca, Piazza della Scienza 3, I-20126 Milano, Italy}
 \affiliation{INFN Milano-Bicocca, Piazza della Scienza 3, I-20126 Milano, Italy}

\author{Michele Pepe}
 \affiliation{INFN Milano-Bicocca, Piazza della Scienza 3, I-20126 Milano, Italy}

\date{\today}

\begin{abstract}
We construct the traceless components of the energy-momentum tensor on the lattice for the two-dimensional $O(3)$ non-linear $\sigma$-model,
such that their correlation functions satisfy the appropriate Ward identities in the continuum limit. The theory is defined by the standard
lattice discretizations of both the action and the energy-momentum tensor. The discretization of space-time entails that the traceless
two-index symmetric irreducible representation of $O(2)$ splits into two irreducible representations of the dihedral group $D_4$, each
requiring an independent renormalization constant. These constants are defined non-perturbatively by imposing discretized versions of
continuum Ward identities for one-point functions in the presence of shifted boundary conditions. The non-perturbative calculations are then
carried out by Monte Carlo simulations, and the resulting renormalization constants are determined with a final accuracy better than a
permille throughout the investigated range of the dimensionless bare coupling constant squared $0\leq g_0^2\leq 0.6$.
\end{abstract}

\maketitle

\section{Introduction}\label{sec:intro}

The two-dimensional $O(3)$ non-linear $\sigma$-model has long been considered as an important theoretical laboratory~\cite{Novikov:1984ac}
for the non-perturbative study of quantum field theories. Although much simpler than four-dimensional non-Abelian gauge theories, it
shares with the Yang-Mills theory several crucial properties like asymptotic freedom, the dynamical generation of a mass-gap, and a rich
non-perturbative infrared dynamics. 

Historically, the model has provided important insights across a broad range of research areas, including 
renormalization-group methods~\cite{Pelissetto:2000ek}, finite-size scaling techniques~\cite{Luscher:1991wu}, topological
effects~\cite{Polyakov:1975yp,Bietenholz:1995zk,deForcrand:2012se,Dunne:2012ae}, lattice discretization 
effects~\cite{Hasenfratz:2000sa,Hasenbusch:2001ht,Balog:2009np,Balog:2009yj,Balog:2012db}, and algorithmic
developments~\cite{Wolff:1988uh,Prokofev:2001ddj,Endres:2006xu,Wolff:2009kp}.
Its integrability~\cite{Zamolodchikov:1977nu} also makes available exact analytical tools, such as the Thermodynamic Bethe Ansatz and
Non-Linear Integral Equations~\cite{Zamolodchikov:1989cf,Destri:1992qk,Destri:1994bv}, to obtain precise analytical predictions 
for several key observables~\cite{Hasenfratz:1990zz,Balog:2003yr}. For these reasons, in lattice field theory, the $O(3)$ model has often
been used as a benchmark for developing analytical and numerical methods before being applied to the more demanding case of
gauge theories in higher dimensions, where numerical simulations are more expensive and systematic effects are often harder to control.

The experience with the $O(3)$ model can prove instrumental for solving problems involving the energy-momentum
tensor $T_{\mu\nu}$ as well, especially for those requiring very expensive numerical simulations in higher dimensional theories.
The range of applications extends from the response of the theory to space-time and scale transformations to the coupling
of the theory to variations of the background metric. At finite temperature, $T_{\mu\nu}$ links the microscopic dynamics
to the macroscopic properties of the thermal theory: the Equation of State (EoS) can be extracted from thermal expectation values
of the various components of $T_{\mu\nu}$ \cite{Landau-hydro}, while transport properties of the medium are
encoded in its correlation functions.

While in the continuum the definition and the conservation of $T_{\mu\nu}$ follow directly from space-time symmetries,
its non-perturbative definition presents significant challenges on the lattice. The hard breaking of continuous
translation invariance requires a non-trivial renormalization of
$T_{\mu\nu}$~\cite{Caracciolo:1989pt,Caracciolo:1989bu,Caracciolo:2000td} before its matrix elements approach the continuum limit with
lattice artifacts vanishing as powers of the lattice spacing up to logarithmic corrections. Progress has been achieved in
Yang-Mills theory~\cite{Giusti:2015daa} and in Quantum Chromodynamics (QCD)~\cite{DallaBrida:2020gux,Bresciani:2026fqc} through the development of
non-perturbative renormalization strategies leading to high-precision determinations of the corresponding renormalization constants.

The aim of this work is to define non-perturbatively the traceless components of the energy-momentum tensor on the lattice for the
two-dimensional $O(3)$ non-linear $\sigma$-model by presenting a non-perturbative determination of their renormalization constants. The key
theoretical novelty is the use of shifted boundary conditions~\cite{Giusti:2010bb,Giusti:2011kt,Giusti:2012yj}, a technique which
has recently been explored in this context in Refs.~\cite{Lauk:2025agt,Lauk:2026jzt}.

This paves the way to compute Euclidean correlation functions of the energy-momentum tensor to address many of the open problems in the
higher dimensional theories. In particular, finite temperature spectral densities and real-time properties associated to $T_{\mu\nu}$, whose
first-principles determination remains one of the major challenges in lattice field theory, can now be investigated similarly to what is
already being done at zero temperature in other channels~\cite{Bulava:2021fre}.

The paper is organized as follows. In Section~\ref{sec:strategy} we review the energy-momentum tensor of the $O(3)$ model
in the continuum and on the lattice, and we present the non-perturbative renormalization strategy. In Section~\ref{sec:numstudy}
we discuss the lattice observables involved in the calculation and the methods employed in the numerical simulations, while the results for
the renormalization constants are discussed in Section~\ref{sec:renconst}. Section~\ref{sec:EoS} is dedicated to the computation of the
Equation of State, while Section~\ref{sec:conclusions} contains our conclusions. The Appendices discuss finite-size effects, present the
perturbative computation useful for improving the definition of the renormalization constants and, finally, collect some tables with the
numerical data obtained from Monte Carlo simulations.

\section{Renormalization strategy}\label{sec:strategy}

\subsection{The energy-momentum tensor in the continuum}

The two-dimensional $O(3)$ non-linear $\sigma$-model is defined by the Euclidean action
\begin{equation}\label{eq:Scont}
S=\dfrac{1}{2 g_0^2}\int d^2x\, \left[ \partial_\mu\vec{\sigma}(x) \right]^2\, ,
\end{equation}
where $\mu=0,1$, $\vec{\sigma}(x)$ is a three-component real field satisfying the constraint $\vec{\sigma}(x)\cdot \vec{\sigma}(x) = 1$,
and $g_0$ is the dimensionless bare coupling. Although the action contains only a quadratic kinetic term, the non-linear constraint induces
interactions among the field components. The model is asymptotically free and generates a mass-gap dynamically. 

The energy-momentum tensor is defined starting from the variation of the action under a change of the metric and it is given by 
\begin{equation}\label{eq:EMTcont}
T_{\mu\nu}(x) =\! \dfrac{1}{g_0^2}\! \left[ \partial_\mu \vec{\sigma}(x)\! \cdot\! \partial_\nu\vec{\sigma}(x)
\!-\! \dfrac{\delta_{\mu\nu}}{2}  \left[ \partial_\rho \vec{\sigma}(x)\right]^2 \right] .
\end{equation}
This equation gives the classical expression; in the quantum theory, the trace acquires an anomalous contribution through renormalization.
Its trace and traceless parts transform under irreducible representations of the Euclidean rotation group $O(2)$. The trace,
$(T_{00}+T_{11})$, is invariant and it is an $O(2)$ scalar, while the two independent components $T_{01}$ and $(T_{00}-T_{11})$
form a symmetric traceless rank-two tensor, transforming in the spin-two irreducible representation of $O(2)$. Furthermore, the component
$T_{01}$ is the Noether current associated with translation in the spatial direction and plays a central role in the renormalization
strategy described below.

To define the energy-momentum tensor non-perturbatively on the lattice we adopt the framework of shifted boundary
conditions~\cite{Giusti:2010bb,Giusti:2011kt,Giusti:2012yj}, which has proved particularly effective for the Yang-Mills
theory~\cite{Giusti:2014ila,Giusti:2015daa} and QCD~\cite{DallaBrida:2020gux,Bresciani:2026fqc}. The system is formulated in a finite
volume of extent $L_0$ in the compact direction and spatial size $L_1$. Along the compact direction the field satisfies the shifted boundary
condition 
\begin{equation}\label{eq:shiftBC}
\vec{\sigma}(x_0+L_0,x_1) = \vec{\sigma}(x_0,x_1-L_0\xi)\, ,
\end{equation}
where $\xi$ is a dimensionless shift parameter, while periodic boundary conditions are imposed in the spatial direction.
The partition function and the corresponding free-energy density are
\begin{equation}\label{eq:Zpart}
{\cal Z} (L_0,\xi) = \int {\cal D}\vec{\sigma}\,\prod_x \delta \left(\vec{\sigma}(x)^2-1)\right) \;  e^{-S},
\end{equation}
\begin{equation}\label{eq:fd}
f(L_0,\xi) = -\dfrac{1}{L_0 L_1}\ln {\cal Z} (L_0,\xi)\, .
\end{equation}
In the thermodynamic limit, Euclidean $O(2)$ invariance implies that the free-energy density depends only on the invariant length of the
compact direction, 
\begin{equation}
f(L_0,\xi) = f(L_0\sqrt{1+\xi^2},0)\, .
\end{equation}
Thus, as already observed in Yang--Mills theory and QCD, a system with shifted boundary conditions is equivalent, in the thermodynamic
limit, to a periodic system with a longer extension in the compact direction $L_0/\gamma$, where $\gamma^{-1}=\sqrt{1+\xi^2}$, and
corresponding to an effective temperature $T=\gamma/L_0$.  

An important consequence of shifted boundary conditions is the equation
\begin{equation}\label{eq:WI_T01}
  \braket{ T_{01} }_\xi =  - \dfrac{\partial}{\partial \xi} f (L_0,\xi)\,,
\end{equation}
which relates the expectation value of the off-diagonal component of the energy-momentum tensor to the response of the free-energy density under an
infinitesimal variation of the shift parameter. Moreover, an interesting Ward identity connects the two components of the traceless tensor
\begin{equation}\label{eq:WI_T01_T00T11}
(1-\xi^2) \braket{ T_{01} }_\xi = \xi  \braket{ T_{00}-T_{11}}_\xi\, .
\end{equation}

These identities are the direct analogue of those employed in the Yang-Mills theory~\cite{Giusti:2014ila,Giusti:2015daa} and in
QCD~\cite{DallaBrida:2020gux,Bresciani:2026fqc} to renormalize non-perturbatively the energy-momentum tensor. After discretization, they are
violated by lattice artifacts and are recovered in the continuum limit provided that the lattice operators are properly renormalized. Thus,
once discretized, they provide a set of non-perturbative renormalization conditions for determining the finite renormalization constants of
the traceless components of the energy-momentum tensor.  

\subsection{The lattice definition and the renormalization conditions}
On the lattice, the continuum $O(2)$ rotation group
is broken to the dihedral subgroup $D_4$ and the traceless two-index symmetric representation splits
into two irreducible representations of $D_4$. Hence, at finite lattice spacing, the energy-momentum
tensor must be reconstructed as a linear combination of all $O(3)$-invariant operators of dimension $d\leq 2$
which transform under those representations. In addition to the identity, there are
only three such operators (no summation over repeated $\mu$ and $\nu$ here)~\cite{Buonanno:1995us,Caracciolo:2000td}: 
\begin{align}\label{eq:EMT_lattop}
   & T_{\mu\nu}^{[1]}(x) = (1 - \delta_{\mu\nu}) \dfrac{1}{g_0^2} 
     \left[   \nabla^{\rm s}_\mu\vec{\sigma}(x) \cdot \nabla^{\rm s}_\nu\vec{\sigma}(x) \right] \, , \nonumber \\
   & T_{\mu\nu}^{[2]}(x) = \dfrac{\delta_{\mu\nu}}{g_0^2} 
        \left[ \nabla^{\rm s}_\rho\vec{\sigma}(x) \right]^2\, , \\
   & T_{\mu\nu}^{[3]}(x) = \dfrac{\delta_{\mu\nu}}{g_0^2} 
    \left[ \nabla^{\rm s}_\mu\vec{\sigma}(x) \cdot \nabla^{\rm s}_\nu\vec{\sigma}(x) 
        -\frac{1}{2} \left[ \nabla^{\rm s}_\rho\vec{\sigma} (x) \right]^2 \right], \nonumber
\end{align}
where $\nabla_\mu^{\rm s} \vec{\sigma}(x)=[\vec{\sigma}(x+a\hat\mu)-\vec{\sigma}(x-a\hat\mu)]/(2a)$ denotes the symmetric
derivative in the $\hat \mu$ direction. The off-diagonal field, $T_{\mu\nu}^{[1]}$, and the traceless diagonal one, $T_{\mu\nu}^{[3]}$,
renormalize multiplicatively~\cite{Caracciolo:2000td}, while the singlet, $T_{\mu\nu}^{[2]}(x)$ represents the trace and mixes also with
the identity operator.
The renormalized energy-momentum tensor which satisfies the continuum Ward identities in the continuum limit can be
written as
\begin{equation}
\hspace{-0.25cm}    T_{\mu\nu}^{\rm R}\!=\! Z_{_T} \left\{ T_{\mu\nu}^{[1]}  + z_{_T} T_{\mu\nu}^{[3]} +
      z_{\rm s} \left[ T_{\mu\nu}^{[2]} - \braket{ T_{\mu\nu}^{[2]} } \right]\right\}.
\end{equation}
where the renormalization constants $Z_{_T}$, $z_{_T}$ and $z_{\rm s}$ are fixed non-perturbatively by requiring that
$T_{\mu\nu}^{\rm R}$ satisfies a discretized version of the continuum Ward identities up to discretization effects
that vanish in the continuum limit. 

In this work we focus on the renormalization of the traceless components of the energy-momentum tensor, namely on the determination of the
renormalization constants $Z_{_T}$ and $z_{_T}$. The former follows from imposing the equation relating the expectation
value of the off-diagonal component of the
energy-momentum tensor to the derivative of the free-energy density with respect to the shift
\begin{equation}\label{ren_cond_ZT}
Z_{_T}(g_0^2, a/L_0) = - \dfrac{\Delta f(L_0,\xi)}{\Delta \xi}\, \dfrac{1}{\braket{ T^{[1]}_{01} }_\xi }\; , 
\end{equation}
where the derivative is discretized by the symmetric finite difference 
\begin{equation}\label{eq:discrder}
\dfrac{\Delta f(L_0,\xi)}{\Delta \xi} = \dfrac{1}{2a L_1}
\ln\Big[\dfrac{{\cal Z} (L_0,\xi_{_-})}{{\cal Z} (L_0,\xi_{_+})} \Big] \; ,
\end{equation}
with $\xi_{_\pm} = \xi \pm a/L_0$.
Similarly, the renormalization constant $z_{_T}$ is obtained by imposing the Ward identity relating the off-diagonal
and the traceless
diagonal components
\begin{equation}\label{ren_cond_zT}
z_{_T} (g_0^2, a/L_0) = \dfrac{1-\xi^2}{\xi} \dfrac{\braket{ T_{01}^{[1]} }_\xi}{\braket{ T_{00}^{[3]} - T_{11}^{[3]} }_\xi} \, .
\end{equation}
In the thermodynamic limit, we remark that $Z_{_T}(g_0^2,a/L_0)$ has discretization effects depending on $(a/L_0)$ of the form
\begin{equation}\label{latart}
  \begin{split}
   & Z_{_T}(g_0^2,a/L_0) = Z_{_T}(g_0^2)  + \\
    \left(\dfrac{a}{L_0}\right)
 & \left[ Z_{_{T,1}} \left( \dfrac{a}{L_0} \right) + Z_{_{T,2}} \left( a\,\Lambda_{O(3)} \right) \right]\! +O(a^4)
\end{split}
\end{equation}
where $Z_{_{T,1}}$ and $Z_{_{T,2}}$ are numerical coefficients; a similar result holds for $z_{_T} (g_0^2, a/L_0)$. These contributions
can be removed by
defining the renormalization constants through the limit 
\begin{align}\label{renZz}
   Z_{_T}(g_0^2) &= \lim_{1/L_0\rightarrow 0} Z_{_T}(g_0^2, a/L_0)\,, \\
   z_{_T}(g_0^2) &= \lim_{1/L_0\rightarrow 0} z_{_T}(g_0^2, a/L_0)\,, \label{renz}
\end{align}
which is taken at a fixed value of the bare coupling $g_0^2$
and shift parameter $\xi$.
This removes the $L_0$-dependent discretization effects as well as the dependence on $\xi$. $L_0$-independent
discretization effects proportional to $(a\, \Lambda_{O(3)})^p$, with $p \geq 2$, remain as part of the definition of
the renormalization constants. Finally, as in Yang-Mills theory and QCD, both renormalization conditions involve
only one-point functions of the fields entering the definition of the energy-momentum tensor. This makes them
particularly suitable for a precise numerical determination of the renormalization constants. 

\section{The numerical study}\label{sec:numstudy}

\subsection{The lattice setup}
\begin{figure}[htb]
  \center
  \includegraphics[width=.8\linewidth]{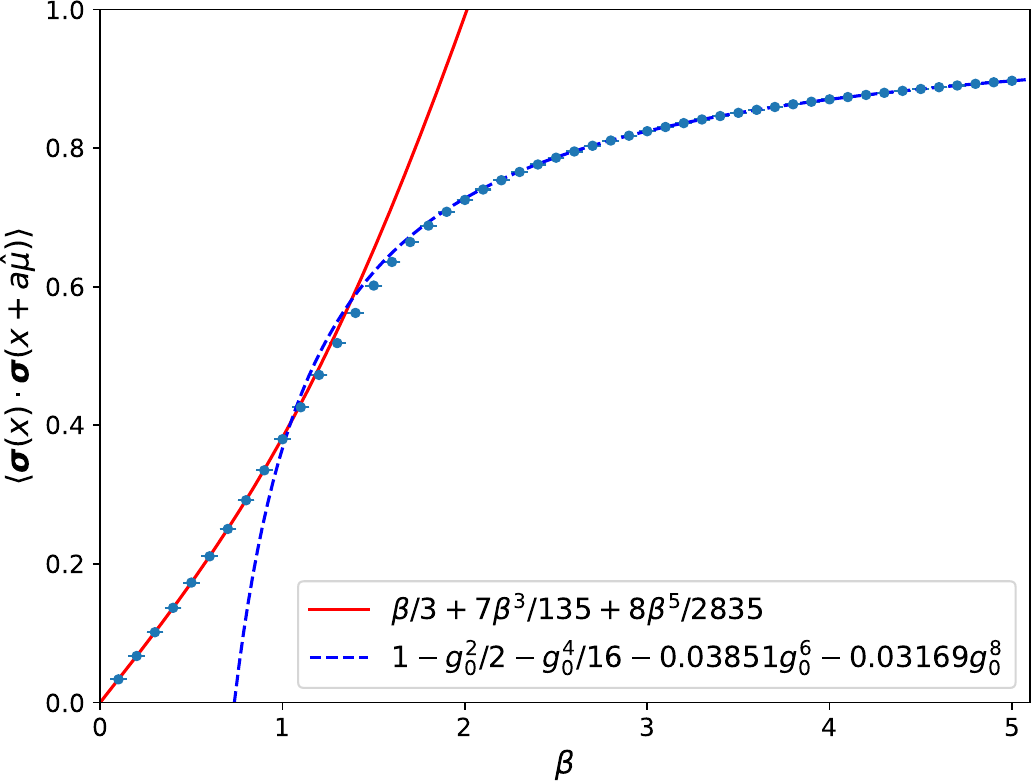}
  \caption{Dependence of $\braket{ \vec{\sigma}(x)\cdot \vec{\sigma}(x+a\hat{\mu}) }$ on the coupling
    $\beta$ for the standard lattice action. The dashed and the continuous curves are the strong~\cite{Campostrini:1996ei} and the
    weak~\cite{Berg:1981er,Alles:1997rr} coupling expansions at $5^{th}$ and $4^{th}$ order, respectively.}
  \label{fig:action_density}
\end{figure}

The action of a quantum field theory regularized on the lattice can be chosen with some freedom and several alternatives have been
considered for the two-dimensional $O(3)$ non-linear $\sigma$-model~\cite{Symanzik:1983dc,Symanzik:1983gh,Hasenfratz:1993sp,Bietenholz:2010xg} with
the purpose of removing lattice artifacts which, in some cases, turned out to be remarkably
large~\cite{Hasenfratz:2000sa,Hasenbusch:2001ht,Balog:2009np,Balog:2009yj,Balog:2012db}. In this study we consider the standard lattice
action, defined as
\begin{equation}\label{eq:latt_act}
    S=-\beta \sum_{x,\mu}~\vec{\sigma}(x)\cdot \vec{\sigma}(x+a\hat{\mu}),
\end{equation}
where $a$ is the lattice spacing, $\hat{\mu}$ is a unit vector in the $\mu$ direction and $\beta=1/g_0^2$ is the bare coupling. We impose
shifted boundary conditions with shift parameter $\xi$ along the compact direction of size $L_0$ as in Eq.~(\ref{eq:shiftBC}), while the
system is periodic in space with extent $L_1$. The standard lattice action has a smooth crossover from a strong coupling regime at small
$\beta$ to the weak coupling limit at large $\beta$ -- i.e. small $g_0^2$ -- where the continuum limit is eventually approached. For
completeness, Fig.~\ref{fig:action_density} shows the known dependence~\cite{Berg:1981er} on the bare coupling of the nearest-neighbor
two-point correlation function, $\braket{ \vec{\sigma}(x)\cdot   \vec{\sigma}(x+a\hat{\mu}) }$, computed in a periodic box of size
$100^2$. For comparison, both the strong~\cite{Campostrini:1996ei} and the weak~\cite{Berg:1981er,Alles:1997rr} coupling expansions at
$5^{th}$ and $4^{th}$ order are plotted, respectively. Since our goal is the non-perturbative determination of renormalization constants
relevant for continuum applications, we restrict our numerical investigation to $\beta \gtrsim 1.65$, corresponding to coupling values in
the range $0 \le g_0^2 \le 0.6$.    

Monte Carlo simulations of the $O(3)$ model can be performed very efficiently with the Wolff cluster algorithm~\cite{Wolff:1988uh}, which
removes critical slowing down and allows for the construction of improved estimators for both local and extended
observables~\cite{Wolff:1989hv}. In the present work, however, all renormalization conditions involve only one-point functions of local
fields. For these observables, a local heat-bath algorithm combined with standard estimators proved to be computationally more efficient,
especially in the weak-coupling regime where cluster updates generate large clusters. We therefore performed all simulations with the local
heat-bath algorithm. 

\subsection{Computation of $\braket{ T_{\mu\nu}^{[1]} }_\xi $ and $\braket{ T_{\mu\nu}^{[3]} }_\xi$}

The bare expectation values $\langle T_{\mu\nu}^{[1]} \rangle _\xi $ and $\braket{ T_{\mu\nu}^{[3]} }_\xi $ have been computed on lattices
with fixed aspect ratio $L_1/L_0=50$ and considering $L_0/a = 6, 8, 10, 12, 14, 20$. Monte Carlo simulations have been carried out at
equally spaced values of the bare coupling in the range $0 < g_0^2 \le 0.6$ in steps of 0.025.

The renormalization constant $Z_{T}$ is determined from Eq.~(\ref{ren_cond_ZT}) using the shift $\xi=1$. Previous studies in Yang-Mills
theory and QCD have shown that lattice artifacts are particularly small for this choice, a feature that is confirmed to hold also in the
$O(3)$ model. The same value for the shift cannot be employed for the determination of $z_{_T}$ because Eq.~(\ref{ren_cond_zT}) becomes singular at
$\xi=1$. We therefore carried out a tree-level perturbative study of lattice artifacts for several integer and half-integer values of the
shift. Although larger values of $\xi$ reduce discretization effects, they also decrease the expectation values of $T_{\mu\nu}^{[1]}$ and
$T_{\mu\nu}^{[3]}$, thereby increasing the statistical uncertainty. The choice $\xi=2$ provides a satisfactory compromise between these
competing effects and is adopted throughout the determination of $z_{_T}$. 

\subsection{Computation of $\dfrac{\Delta f (L_0,\xi)}{\Delta \xi}$ }\label{Df/Dxi}
The determination of the renormalization constant $Z_{_T}$ using Eq.~(\ref{ren_cond_ZT}) requires the computation of the discrete derivative of
the free-energy density with respect to the shift parameter. Employing a symmetric definition, this derivative can be expressed as the ratio
of the two partition functions given in Eq.~(\ref{eq:discrder}). We have computed this ratio at shift parameter $\xi =1$ using the bare
couplings and the lattice sizes described in the previous subsection.

The numerical evaluation of a ratio of partition functions is notoriously difficult because the corresponding relevant phase spaces
generally have a very small overlap. In Yang-Mills theory and QCD this problem has been successfully addressed by expressing the ratio as an
integral over the bare coupling of the difference between the corresponding action
densities~\cite{Giusti:2015daa,Bresciani:2025vxw,Bresciani:2026fqc}. Although this approach is fully applicable also in the present case, we
use instead the interpolation method proposed in Ref.~\cite{deForcrand:2000fi}. It introduces a sequence of intermediate systems
connecting the two theories of interest, so that successive partition functions have a substantial overlap and their ratios can be estimated
efficiently. While the computational cost grows linearly with the spatial volume, this method has the advantage that the partition function
ratio is obtained independently at each value of the bare coupling. Owing to the modest computational cost of the two-dimensional $O(3)$
model, this strategy is particularly convenient and was therefore adopted throughout the present work. 

We therefore define a new $O(3)$ system with the same lattice action as Eq.~(\ref{eq:latt_act}) but such that, when coupling the spins across the
compact direction, we consider 
\begin{equation}
\vec\sigma(L_0,x_1) = \lambda\,  \vec\sigma\left(0,x_1^{\tiny -}\right) + (1-\lambda)\, \vec\sigma\left(0,x_1^{\tiny +} \right) \, .
\end{equation}
where $x_1^{\tiny \pm} = x_1-\xi_{_\pm} L_0$.
In the spatial direction we have usual periodic boundary conditions and we indicate with ${\cal Z}_\lambda(L_0,\xi)$ the corresponding partition
function. We notice that the partition functions at the numerator and at the denominator of Eq.~(\ref{eq:discrder}) correspond to $\lambda=1$
and $\lambda=0$, respectively. We now introduce ${\cal N}\!-\!1$ interpolating systems and we rewrite the ratio of the two initial partition
functions as 
\begin{align}\label{chainZ}
  \dfrac{{\cal Z}_1}{{\cal Z}_0}=\dfrac{{\cal Z}_\epsilon}{{\cal Z}_0} \cdot&
  \dfrac{{\cal Z}_{2\epsilon}}{{\cal Z}_\epsilon} \cdot \ldots \cdot \dfrac{{\cal Z}_1}{{\cal Z}_{1-\epsilon}} =
  \prod_{n=0}^{{\cal N}-1} \dfrac{{\cal Z}_{(n+1)\epsilon}}{{\cal Z}_{n \epsilon}}\, , \\
  &\log \dfrac{{\cal Z}_1}{{\cal Z}_0} = \sum_{n=0}^{{\cal N}-1} \log \dfrac{{\cal Z}_{(n+1)\epsilon}}{{\cal Z}_{n \epsilon}}\, ,
\end{align}
where $\epsilon=1/{\cal N}$. If $\epsilon$ is sufficiently small, successive partition functions have relevant phase space with a large overlap so
that it is possible to estimate their ratio efficiently by a single Monte Carlo importance sampling. More explicitly, introducing the action
corresponding to the midpoint value $\lambda=(n+1/2)\epsilon$, the ratio of two successive partition functions can be rewritten as a ratio
of expectation values evaluated in the same ensemble
\begin{equation}\label{chainZ_obs}
  \dfrac{{\cal Z}_{{(n+1)\epsilon}}}{{\cal Z}_{{n \epsilon}}} =
  \dfrac
  {\braket{ e^{-\epsilon\, \Delta S/2} }_{_{(n+1/2)\epsilon}}}
  {\braket{ e^{+\epsilon\, \Delta S/2} }_{_{(n+1/2)\epsilon}}} \, ,
\end{equation}
where the expectation values on the r.h.s. are computed in the system with partition function $Z_{(n+1/2)\epsilon}$ and the variation of the
action is
\begin{equation}\label{DeltaS}
\hspace{0.125cm}  \Delta S\!=\! \beta\!\! \sum_{x_1=0}^{L_1-a}\!\!\! \vec \sigma(L_0\!- a,x_1)\! 
  \Big[ \vec\sigma\!\left(0, x_1^{\tiny +}\right) - \vec\sigma\!\left(0, x_1^{\tiny -}\right)\! \Big] .
\end{equation}
Efficient importance sampling requires the reweighting factors appearing in Eq.~(\ref{chainZ_obs}) to be close to 1. In practice this
is achieved by choosing the number of interpolating systems such that $| \epsilon\, \Delta S/2| \lesssim 1$, which implies approximately
$\epsilon L_1/a\simeq1$. Accordingly, we used ${\cal N}=500$, 800, 1000, 1250, 1600 and 2000 for $L_0/a=6$, 8, 10, 12, 14 and 20, respectively.

\section{Determination of the renormalization constants}\label{sec:renconst}

The renormalization constants are determined following the strategy described in Section~\ref{sec:strategy}. For each value of the bare
coupling, Monte Carlo simulations were performed on lattices with $L_0/a= 6$, 8, 10, 12, 14 and 20, and shift parameter $\xi=1$
for $Z_{_T}(g_0^2, a/L_0)$ and $\xi=2$ for $z_{_T}(g_0^2, a/L_0)$. The spatial volume is chosen by fixing the aspect ratio to $L_1/L_0=50$. A
dedicated study of the dependence on the spatial volume, presented in Appendix~\ref{FVeffects}, shows that residual spatial finite-volume effects are
negligible compared to the statistical uncertainties of the present calculation. We have also measured the topological charge for $L_0/a=6$
using the geometric definition. The Monte Carlo histories show tunneling between distinct topological sectors, with a probability that
decreases as $g_0^2$ becomes smaller. For $g_0^2\lesssim 0.3$, tunneling events become strongly suppressed and the system remains in
the trivial sector as expected, see Refs.~\cite{DallaBrida:2021ddx,Bresciani:2025mcu,Bresciani:2026fqc}. 

The renormalization constants are obtained by extrapolating the finite-$L_0$ estimates to the $1/L_0\rightarrow0$ limit at fixed bare
coupling. This extrapolation can be improved by removing the leading finite-$L_0$ corrections computed in lattice perturbation theory. To
one-loop order, this amounts to replacing  
\begin{equation}
  X(g_0^2, \dfrac{a}{L_0}) \to
  \dfrac{X(g_0^2, \dfrac{a}{L_0}) \Big[ X^{(0)} + g_0^2\, X^{(1)}\Big]} {\left[ X^{(0)}(\dfrac{a}{L_0}) + g_0^2\, X^{(1)}(\dfrac{a}{L_0})\right] } \, ,
\end{equation}
where $X = Z_{_T}, z_{_T}$ and $X^{(0)}(a/L_0)$ and $X^{(1)}(a/L_0)$ are the tree-level and one-loop coefficients at finite $L_0/a$. Their
calculation is summarized in Appendix~\ref{appPT}, while the corresponding infinite-volume coefficients, $X^{(0)}$ and $X^{(1)}$, in the
$1/L_0\rightarrow 0$ limit are taken from~\cite{Caracciolo:2000td}.  

\subsection{Renormalization constant of $T_{\mu\nu}^{[1]}$}

The renormalization constant $Z_{_T}$ is determined using Eqs.~(\ref{ren_cond_ZT}) and~(\ref{renZz}). The Monte Carlo simulations were
performed with shift $\xi=1$ at equally spaced values of the bare coupling within the range $0 < g_0^2 \le 0.6$. The numerical results are
collected in Tab.~\ref{tab:T01_xi1} and Tab.~\ref{tab:dfdxi} in Appendix~\ref{appTables}. In the top panel of Fig.~\ref{fig:ZT_cont_all}, we
plot the $a/L_0$ dependence of $Z_{_T}(g_0^2, a/L_0)$ for the investigated values of $g_0^2$. The dependence on $a/L_0$ is remarkably weak 
over the whole range of bare couplings investigated. Consequently, the one-loop perturbative improvement has only a marginal numerical
impact. Within our statistical precision no linear dependence on $a/L_0$ can be resolved and the extrapolations are therefore performed
using a quadratic ansatz. This behaviour is consistent with the expectation that the coefficient of the linear term is suppressed by the
non-perturbative scale $(a\,\Lambda_{O(3)})$ which, for instance, rapidly drops from $O(0.01)$ to $O(0.001)$ as $g_0^2$ decreases from
$\sim 0.6$ to $\sim 0.5$, decreasing even more sharply at smaller bare couplings. The extrapolations are carried out independently at the
various values of $g_0^2$ with a sub-permille final accuracy. All fits turn out to be of good quality, with a $\chi^2$/dof ranging from
$0.2$ to $1.8$, as the figure shows.  

\begin{figure}[htb]
  \center
  \includegraphics[width=1\linewidth]{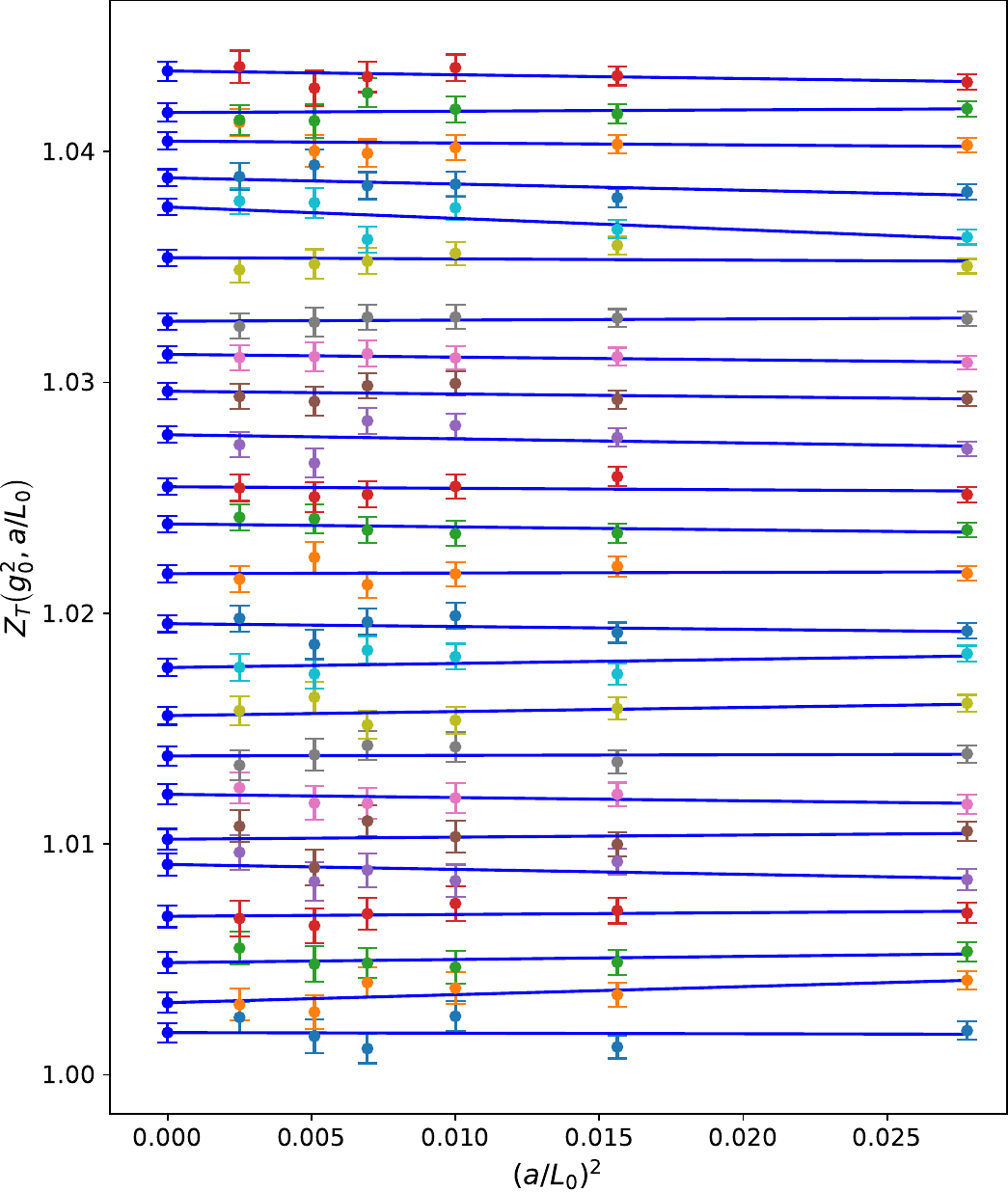}\\
  \includegraphics[width=1\linewidth]{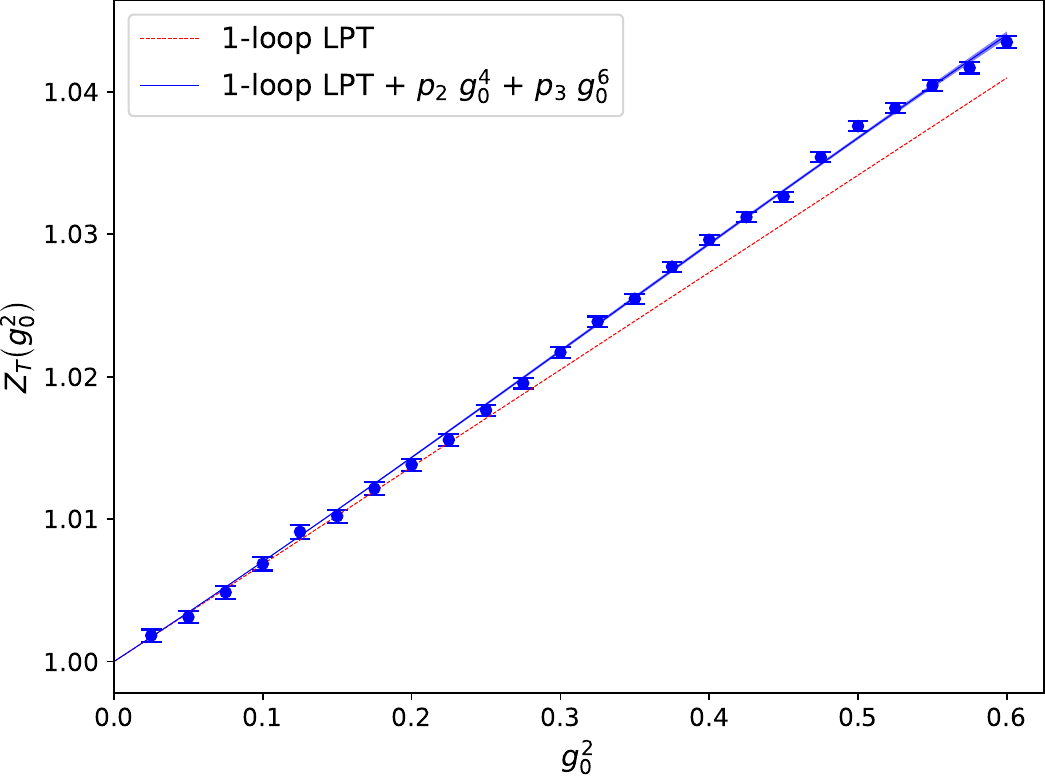}
  \caption{
(Top) Dependence of $Z_{_T} (g_0^2, a/L_0)$ on $a/L_0$ along with the corresponding $1/L_0\rightarrow 0$ extrapolations. The datasets correspond to
increasing values of $g_0^2$ from bottom to top.
(Bottom) Dependence of the renormalization constant $Z_{_T}(g_0^2)$ on the bare coupling. The red and the blue curves represent the
one-loop perturbative prediction and a polynomial fit of order 3, respectively.} 
\label{fig:ZT_cont_all}
\end{figure}

The extrapolated values $Z_{_T}(g_0^2)$ are listed in Tab.~\ref{tab:Zs} and shown in the bottom panel of Fig.~\ref{fig:ZT_cont_all} as a
function of $g_0^2$. The one-loop perturbative prediction is represented by the red line, while the blue curve corresponds to a fit of the form
\begin{equation}
  Z_{_T} (g_0^2) = Z_{_T}^{(0)} + Z_{_T}^{(1)}\, g_0^2 + p_2 \, g_0^4 + p_3 \, g_0^6 \, ,
\end{equation}
where we enforced the perturbative values $Z_T^{(0)}=1$ and $Z_T^{(1)}=1/\pi-1/4$. The estimated values of the free fitting parameters are
$p_2 = 0.0210(27)$ and $p_3 = -0.021(5)$, with covariance $-1.44\times10^{-5}$ and a resulting $\chi^2/{\rm dof} = 1.08$. 

We observe that the non-perturbative results differ from the one-loop perturbative prediction by at most a few parts per thousand over the
entire range of bare couplings investigated. The largest deviations occur towards the strong-coupling end of the explored interval. Although
small in absolute size, these effects are statistically significant owing to the high numerical precision achieved in the calculation. 

\subsection{Renormalization constant of $T_{\mu\nu}^{[3]}$}
The renormalization constant of the traceless diagonal component is obtained as the product of the overall constant $Z_{_T}$ computed
in the previous subsection, and the relative renormalization constant $z_{_T}$ that we now determine from Eqs.~(\ref{ren_cond_zT})
and~(\ref{renz}).  

We performed Monte Carlo simulations at $L_0/a = 6, 8, 10, 12, 14, 20$ with shift $\xi=2$ computing the expectation values
$\braket{ T_{01}^{[1]} }_{\xi=2}$ and $\braket{ T_{00}^{[3]} - T_{11}^{[3]} }_{\xi=2}$. The numerical data are
tabulated in Tab.~\ref{tab:T01_xi2} and in Tab.~\ref{tab:T00_T11_xi2} in Appendix~\ref{appTables}.
The top panel of Fig.~\ref{fig:zT_trp_imp_cont_all} shows the one-loop improved estimates of $z_{_T}(g_0^2,a/L_0)$. In contrast to the case
of $Z_{_T}$, a residual dependence on $a/L_0$ is clearly visible, although it remains mild over the whole range of couplings. The
perturbative improvement removes a significant fraction of these effects, leaving only small corrections that are eliminated
through the $1/L_0\rightarrow 0$ extrapolation performed independently at each value of the bare coupling. 

The extrapolations are performed using the data obtained at $L_0/a = 8, 10, 12, 14, 20$, with a quadratic ansatz.
Within the statistical precision of the data, no significant linear contribution is observed, in agreement with the expectation
discussed in the previous subsection. As a consistency check, we repeat the fits including the $L_0/a=6$ data. 

The extrapolations yield statistically good quality fits over the full range of bare couplings considered, with a
$\chi^2$/dof ranging from 0.07 to 2.3 and attaining a final accuracy below 0.1~\permil.
The resulting values of $z_{_T}(g_0^2)$ are listed in Tab.~\ref{tab:Zs} and displayed in the bottom panel of
Fig.~\ref{fig:zT_trp_imp_cont_all} as a function of the bare coupling. The dashed red line represents the one-loop perturbative 
expectation, while the continuous blue curve is obtained by fitting the extrapolated values to the polynomial expression
\begin{equation}
  z_{_T} (g_0^2) = z_{_T}^{(0)} + z_{_T}^{(1)}\, g_0^2 + q_2 \, g_0^4 + q_3 \, g_0^6 + q_4 \, g_0^8 \, ,
\end{equation}
where the first two coefficients are constrained to their perturbative values, $z_T^{(0)}=1$ and $z_T^{(1)}=(1-3/\pi)/2$. Treating the higher order
coefficients as free parameters, the fit yields the estimates $q_2 = -0.0030(18)$, $q_3 = 0.025(8)$ and $q_4 = -0.054(9)$, with a
covariance given by ${\rm cov}(q_2, q_3) = -1.51 \times 10^{-5}$, ${\rm cov}(q_2, q_4) = 1.60 \times 10^{-5}$ and
${\rm cov}(q_3, q_4) = -7.45 \times 10^{-5}$, and a resulting $\chi^2/{\rm dof} = 0.44$. 

The renormalization constant remains remarkably close to 1 throughout the investigated range, differing from unity by at most about
1\%. The non-perturbative results are consistent with the one-loop perturbative prediction within statistical uncertainties up to
approximately $g_0^2\simeq0.3$. At larger couplings, deviations reaching approximately 3 parts per thousand become visible. Although
quantitatively very small, these differences are statistically significant thanks to the high numerical precision attainable in the 
two-dimensional $O(3)$ model. 
\begin{figure}[htb]
    \center
    \includegraphics[width=1\linewidth]{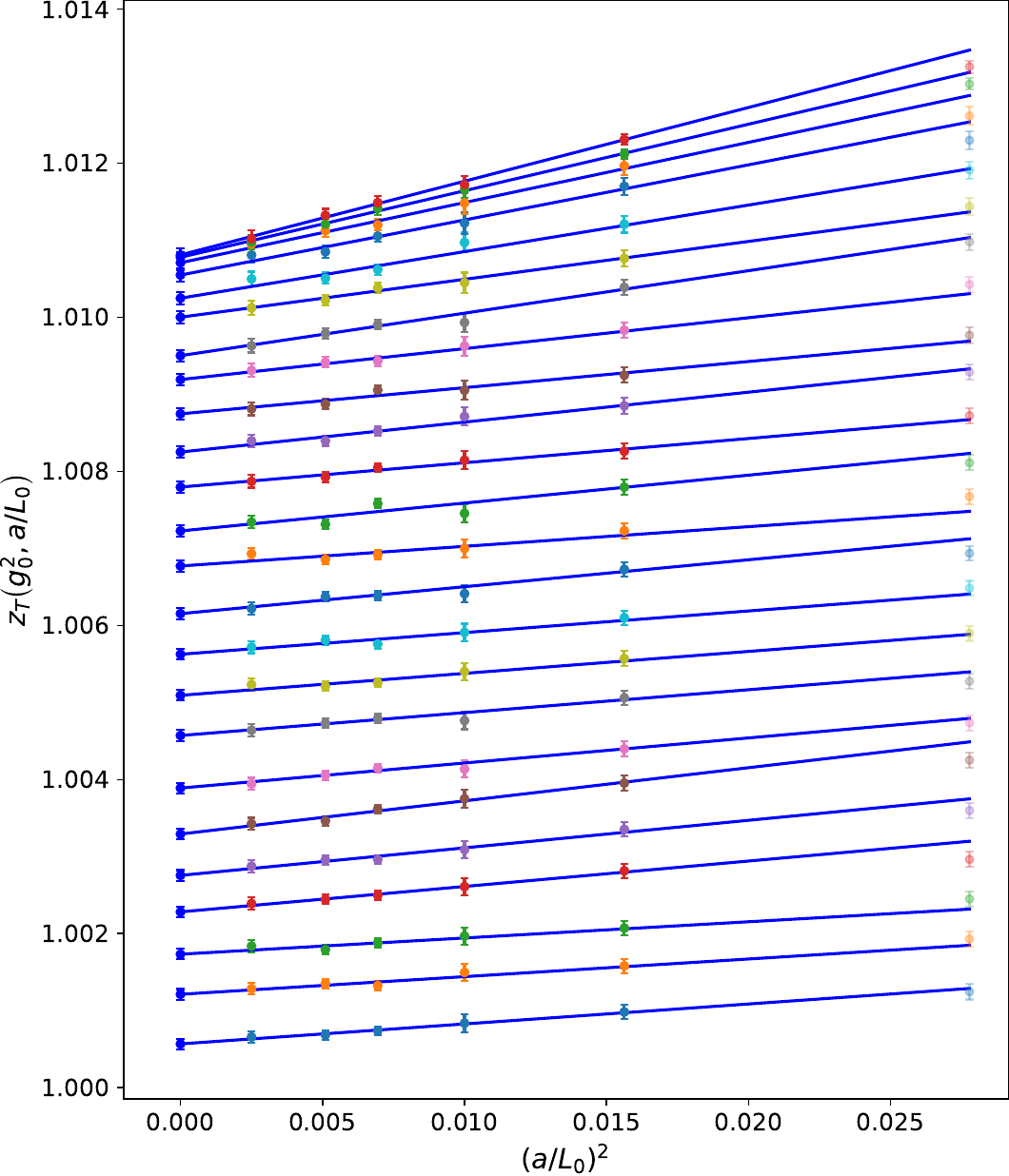}\\
    \includegraphics[width=1\linewidth]{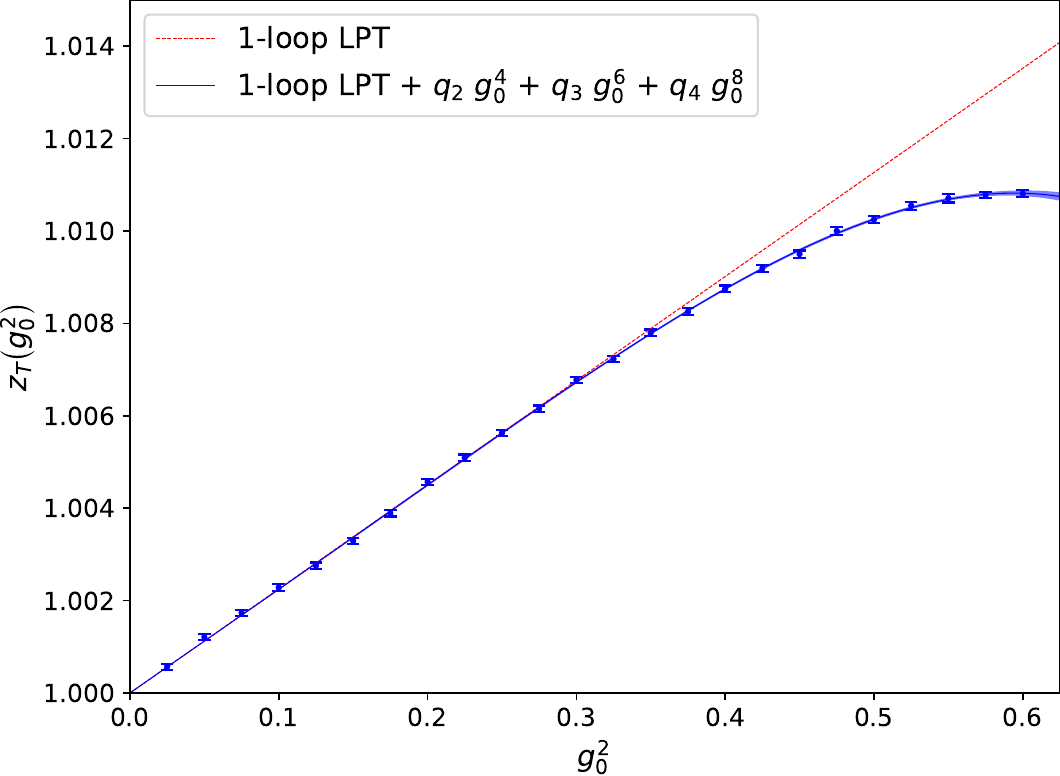}
    \caption{(Top) Dependence of the improved $z_{_T} (g_0^2, a/L_0)$ on $a/L_0$ along with the corresponding $1/L_0\rightarrow 0$
      extrapolations. The  datasets correspond to 
      increasing values of $g_0^2$ from bottom to top.
      (Bottom) Dependence of the renormalization constant $z_{_T}(g_0^2)$ on the bare coupling. The red dashed line and the
      blue continuous curve represent the one-loop 
      perturbative prediction and a polynomial fit of order 4, respectively.}  
    \label{fig:zT_trp_imp_cont_all}
\end{figure}
\begin{table}[ht!]
\centering
\begin{tabular}{|c||c|c|}
\hline
$g_0^2$ & $Z_T$ & $z_T$\\ \hline
 0.025 & 1.0018(4) & 1.00057(7) \\
 0.050 & 1.0031(4) & 1.00121(7) \\
 0.075 & 1.0049(5) & 1.00173(7) \\
 0.100 & 1.0069(5) & 1.00228(7) \\
 0.125 & 1.0091(5) & 1.00276(7) \\
 0.150 & 1.0102(5) & 1.00329(7) \\
 0.175 & 1.0122(4) & 1.00389(7) \\
 0.200 & 1.0138(4) & 1.00457(7) \\
 0.225 & 1.0156(4) & 1.00509(7) \\
 0.250 & 1.0176(4) & 1.00563(7) \\
 0.275 & 1.0195(4) & 1.00615(7) \\
 0.300 & 1.0217(4) & 1.00677(7) \\
 0.325 & 1.0239(4) & 1.00723(7) \\
 0.350 & 1.0255(4) & 1.00780(7) \\
 0.375 & 1.0277(4) & 1.00825(7) \\
 0.400 & 1.0296(4) & 1.00875(7) \\
 0.425 & 1.0312(4) & 1.00919(8) \\
 0.450 & 1.0327(4) & 1.00950(8) \\
 0.475 & 1.0354(4) & 1.01000(8) \\
 0.500 & 1.0376(4) & 1.01025(8) \\
 0.525 & 1.0389(4) & 1.01055(9) \\
 0.550 & 1.0405(4) & 1.01071(9) \\
 0.575 & 1.0417(4) & 1.01078(7) \\
 0.600 & 1.0435(4) & 1.01081(8) \\
\hline
\end{tabular}

\caption{The renormalization constants $Z_T (g_0^2)$ and $z_T (g_0^2)$ after the $1/L_0 \to 0$ extrapolations.}   
  \label{tab:Zs}
\end{table}

\section{Equation of State\label{sec:EoS}}
As a test of our determination of $Z_{_T}$, we compute the Equation of State of the two-dimensional $O(3)$ non-linear $\sigma$-model
at a fixed temperature and compare it with the exact prediction obtained from the Thermodynamic Bethe Ansatz and the Non-Linear Integral
Equations of the model. In particular, the ground state energy $E_0 (\mu)$ and the finite-volume mass-gap, $m(\mu)= E_1(\mu) - E_0(\mu)$,
can be determined exactly for a system of finite spatial size $1/\mu$ following Refs.~\cite{Balog:2003yr,Hegedus:2004xd,Balog:2009ze}.
By exchanging the Euclidean temporal and spatial directions, one relates the finite-volume ground state energy to the pressure at temperature
$T=\mu$: $p(T)/T^2 = -E_0(T)/T$. The entropy density, which is the primary thermodynamic observable that can be computed in the
framework of shifted boundary conditions, is then obtained from its definition $s(T)=dp(T)/dT$.

On the lattice, we determine the values of the bare coupling corresponding to constant physics by keeping fixed the value of a
renormalized coupling, following the same strategy employed in the Yang-Mills theory and in QCD. We use the finite-volume coupling
introduced in Ref.~\cite{Luscher:1991wu}, $\bar g^2 (\mu) = m(\mu)/\mu$ and we select the set of parameters corresponding to
$\bar g^2 (\mu) = 0.9176(5)$  (see Table~2 in Ref.~\cite{Luscher:1991wu}) for which the values of the bare coupling $g_0^2$ are in the range
explored in our determination of $Z_{_T}$. From the exact finite-volume mass gap, we obtain $\mu/M = 7.771(24)$, where $M$ is the
infinite-volume mass-gap generated by dimensional transmutation~\cite{Hasenfratz:1990zz}. We then perform Monte Carlo simulations at the
same values of $L_0/a$ and $g_0^2$, imposing shifted boundary conditions with $\xi=1$. The physical temperature, thus, corresponds to
$T=\mu/\sqrt{2}$ and the entropy density is determined from the equation
\begin{equation}\label{eq:ent_dens}
  \dfrac{s(T)}{T} = -\dfrac{L_0^2 (1+\xi^2)^2}{\xi} Z_{_T} \braket{ T_{01}^{[1]} }_\xi \, .
\end{equation}

We fix the aspect ratio to $L_1/L_0=50$ so that, as discussed in Appendix~\ref{FVeffects}, finite-volume effects on $\braket{T_{01}^{[1]}}_{\xi=1}$
are negligible for $g_0^2 \ge 0.35$, as is the case for all the simulations considered here. In Tab.~\ref{tab:EoS} we
collect the results of this computation. 
\begin{table}[h]
  \centering
\begin{tabular}{|c|c||c|}
\hline 
$\beta$ & $L_0/a$ & $\braket{T_{01}^{[1]}}_{\xi=1}\times 10^3$ \\
\hline
1.6952 & 5 & -15.1527(18) \\ 
1.7276 & 6 & -10.5612(16) \\ 
1.7553 & 7 & -7.7813(14) \\ 
1.7791 & 8 & -5.9702(9) \\ 
1.8171 & 10 & -3.8298(5) \\ 
1.8497 & 12 & -2.6659(5) \\ 
1.8965 & 16 & -1.5023(3) \\ 
1.9631 & 24 & -0.66915(14) \\ 
2.0104 & 32 & -0.37688(11) \\ 
\hline
\end{tabular}
\caption{Bare parameters and expectation values $\braket{T_{01}^{[1]}}_{\xi=1}$ corresponding to the temperature $T=\mu/\sqrt{2}$ with
  $\mu/M = 7.771(24)$.}    
  \label{tab:EoS}
\end{table}

\begin{figure}[htb]
    \center
    \includegraphics[width=1\linewidth]{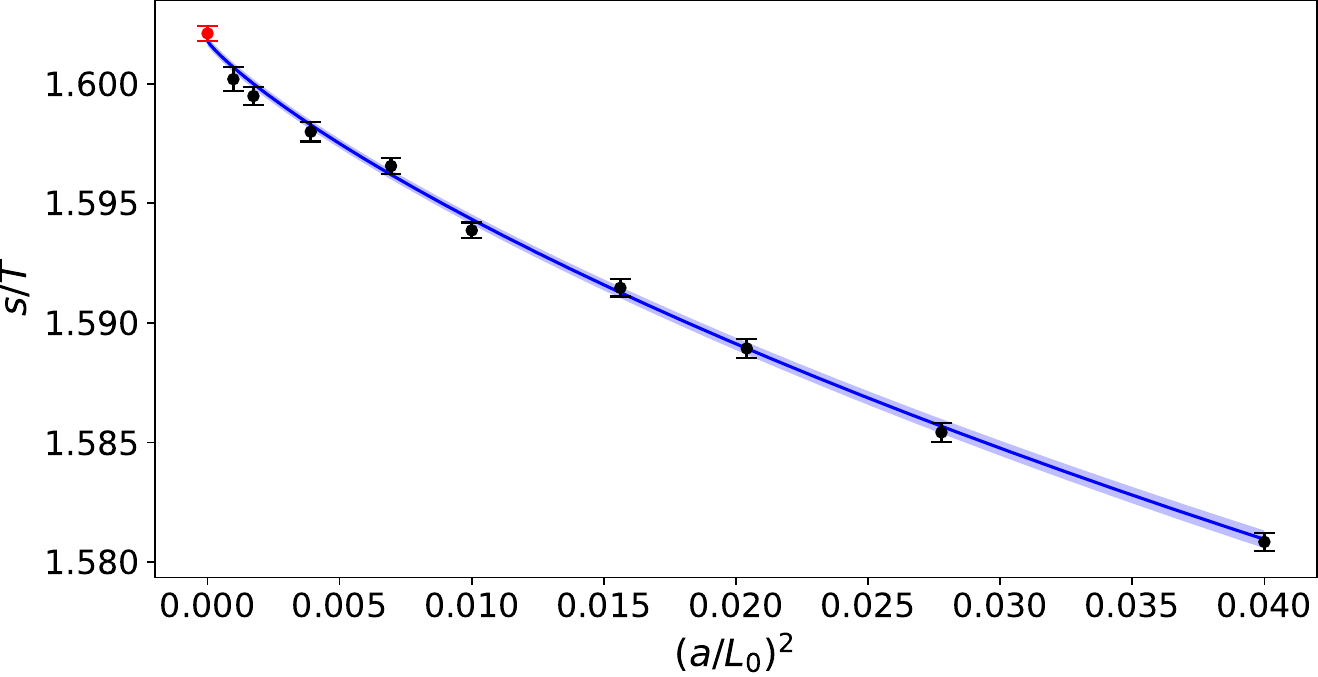}\\
    \caption{Interpolation of the Monte Carlo data for the entropy density $s(T)/T$ (black dots) and of the exact continuum value (red point) at
      the temperature $T=\mu/\sqrt{2}$ with $\mu/M = 7.771(24)$ using the phenomenological fit function given in Eq.~(\ref{eq:sTinterp}).}
    \label{fig:entropy}
\end{figure}

In Fig.~\ref{fig:entropy} we show our numerical results for the entropy density $s(T)/T$ as a function of $(a/L_0)^2$ together with the
analytically computed continuum value $s_0^{\rm cont}=1.6021(3)$. In the two-dimensional $O(3)$ non-linear $\sigma$-model, large
logarithmic corrections affect $O(a^2)$ lattice artifacts~\cite{Hasenfratz:2000sa,Hasenbusch:2001ht,Balog:2009np,Balog:2009yj,Balog:2012db},
mimicking a linear dependence on the lattice spacing. A combined interpolation of our numerical data and of the continuum value using the phenomenological
ansatz 
\begin{equation}\label{eq:sTinterp}
  \dfrac{s(T)}{T}= s_0 + s_1 \left(\dfrac{a}{L_0}\right)^2 \log\left[\left(\dfrac{a}{L_0}\right)^2 \right]\, .
\end{equation}
describes the data within their statistical uncertainties, with $s_0 = 1.6018(2)$, $s_1 = 0.1617(28)$,
cov$(s_0,s_1) = 1.25\cdot 10^{-7}$ and $\chi^2$/dof=1.44. 
The resulting fit is shown as a blue band in Fig.~\ref{fig:entropy}. Fitting the numerical data alone with the same
ansatz yields $s_0 = 1.6015(3)$. The extrapolated value $s_0$ is statistically consistent with the exact continuum
prediction $s_0^{\rm cont}$, with sub-permille precision: this agreement provides a stringent non-perturbative test of our
determination of $Z_T$, which differs from unity by $3$--$4\%$ in the range of bare couplings considered. 

\section{Conclusions}\label{sec:conclusions} 
In this work we have established a fully non-perturbative definition of the traceless components of the energy-momentum tensor in the
two-dimensional $O(3)$ non-linear $\sigma$-model whose correlation functions satisfy the appropriate Ward identities in the continuum limit.
The crucial step forward to reach this goal has been the non-perturbative determination of the necessary renormalization constants. The
calculation is based on Ward identities in a moving reference frame and follows the strategy successfully developed in recent years for the
Yang-Mills theory~\cite{Giusti:2015daa} and QCD~\cite{DallaBrida:2020gux,Bresciani:2026fqc}. By adapting these techniques to the $O(3)$
model, we have obtained, with a sub-permille precision, a non-perturbative determination of the renormalization constants of the traceless
components of the energy-momentum tensor on the lattice as given in Eq.~(\ref{eq:EMT_lattop}) using the standard lattice action over the
range $0 \le g_0^2 \le 0.6$ of the bare coupling.   

The $O(3)$ model -- while sharing many qualitative features with non-Abelian gauge theories -- represents an ideal framework in which to
investigate and validate new non-perturbative methods due to the reduced computational cost of numerical simulations. In particular,
the non-perturbative renormalization of the energy-momentum tensor is a necessary step for future applications, like, for
instance, the developments of new strategies for the computation of spectral densities at finite temperature and transport coefficients, which
are challenging quantities to compute in lattice field theory. Once validated in the simplified framework of the $O(3)$ model, such new
methodologies can be transferred to the Yang-Mills theory and QCD, where they may contribute to improving first-principles determinations of
real-time dynamical observables.

\section{Acknowledgments}\label{sec:ACK}
M.P. thanks Mika Lauk for discussions on the topic of this paper. The numerical simulations have been carried out
on the PC cluster Knuth at Milano-Bicocca, and on the the supercomputer Heian at YITP in Kyoto University.
We thank both these institutions for granting us access to their supercomputers and for the technical support.
This work is (partially) supported by ICSC - Centro Nazionale di Ricerca in High Performance Computing, Big
Data and Quantum Computing, funded by European Union - NextGenerationEU.

\appendix
\section{Finite-volume effects}\label{FVeffects} 
In this Appendix, we present an evaluation of spatial finite-size effects on the observables and on the renormalization constants of the
energy-momentum tensor. In particular, we compare numerical results obtained at aspect ratios $L_1/L_0 = 50$ and $100$ for a fixed extent
$L_0/a = 6$ in the compact direction, where finite-size effects are expected to be largest for a given value of the bare coupling. The
simulations at $L_1/L_0 = 100$ were carried out at the bare coupling values of $g_0^2 = 0.1$, $0.35$, and $0.6$. 

Since the two-dimensional $O(3)$ model has a dynamically generated mass-gap, finite-volume corrections induced by the finite spatial extent
are exponentially suppressed. Denoting by $m(L_0)$ the mass-gap at finite $L_0$, the leading corrections are expected to scale as
$e^{^{-m(L_0/\gamma)\, L_1 \gamma}}$ up to polynomial prefactors. Introducing the renormalized coupling~\cite{Luscher:1991wu}
$g_R^2(L_0/\gamma)=m(L_0/\gamma)L_0/\gamma$, this suppression factor can be rewritten as $e^{^{-g_R^2(L_0/\gamma) \, \gamma^2 L_1/L_0}}$.
As discussed below, finite spatial-volume effects are indeed visible in the expectation values of the energy-momentum tensor, particularly
towards the weak-coupling region where the renormalized coupling becomes smaller. However, these effects largely cancel in the ratios defining
the renormalization constants, so that the finite-volume dependence of $Z_{_T}$ and $z_{_T}$ at aspect ratio $L_1/L_0 = 50$, is negligible
within the statistical accuracy of the numerical data in the whole range of coupling $0 < g_0^2 \le 0.6$.

\subsection{$\braket{T_{01}^{[1]}}_{\xi=1}$ and $\dfrac{\Delta f}{\Delta \xi}$ at $\xi=1$}
\begin{figure}[htb]
    \center
    \includegraphics[width=.8\linewidth]{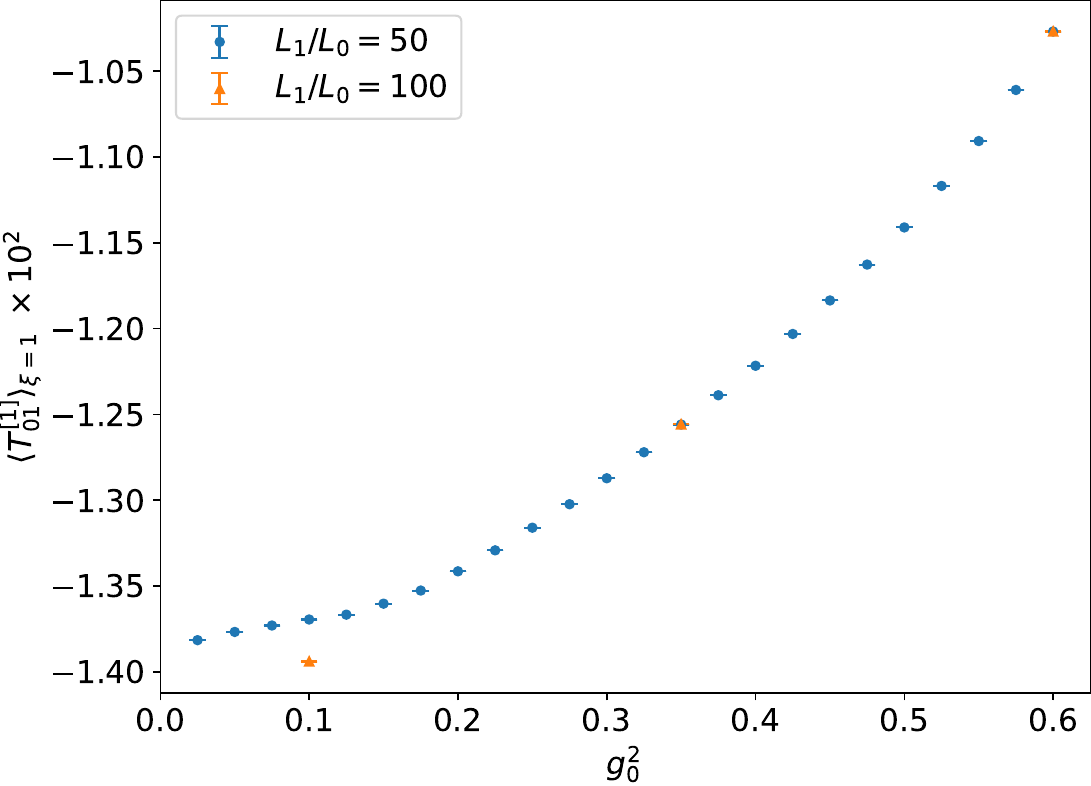}\\
    \includegraphics[width=.8\linewidth]{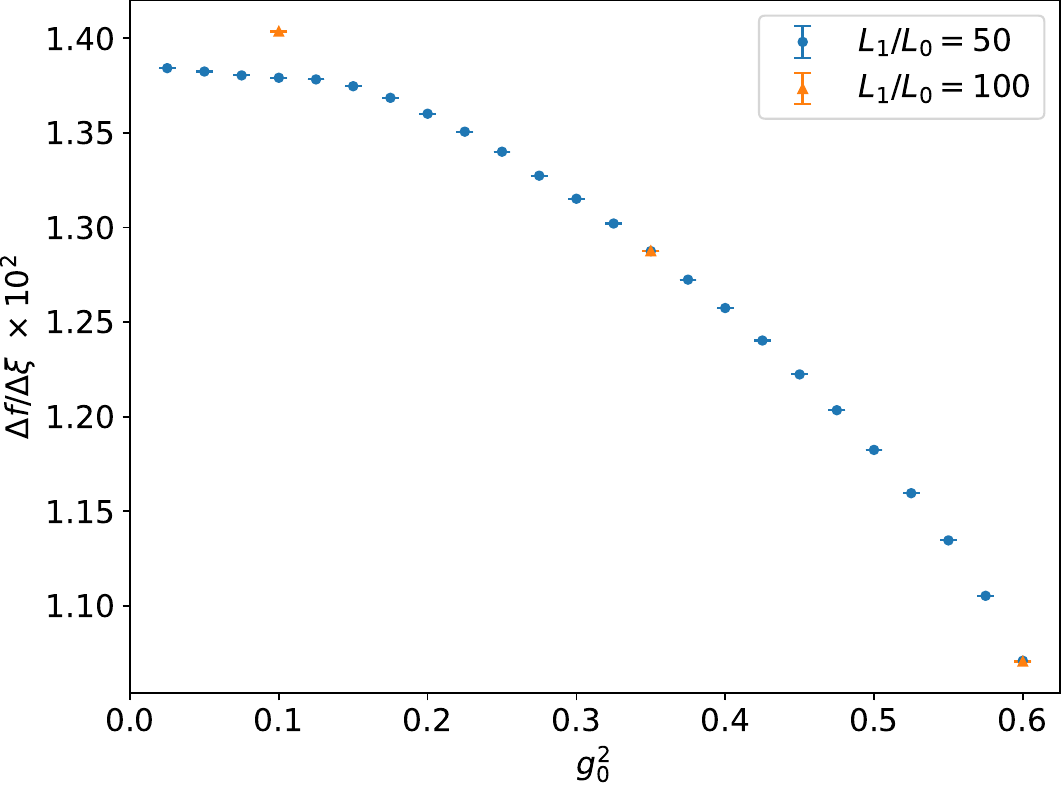}
    \caption{Comparison of the results at $L_1/L_0 = 50$ (blue-circle points) and at 100 (orange-triangle points) for
    $\braket{T_{01}^{[1]}}_{\xi=1}$ (top) and $\Delta f/\Delta\xi$ (bottom) as a function of $g_0^2$ for $L_0/a=6$.}
    \label{fig:finV_T01_dfdxi}
\end{figure}

In the top panel of Fig.~\ref{fig:finV_T01_dfdxi}, we display the bare coupling dependence of the estimate of
$\braket{T_{01}^{[1]}}_{\xi=1}$, evaluated at shift parameter $\xi = 1$. The blue data
points correspond to simulations performed with an aspect ratio of $L_1/L_0 = 50$, while the three orange points represent measurements taken
at $L_1/L_0 = 100$. Based on a statistics of several $10^8$ Monte Carlo sweeps for each value of $g_0^2$, the precision of these
measurements reaches an accuracy of about $0.1\permil$.  

The bottom panel of Fig.~\ref{fig:finV_T01_dfdxi} shows a analogous plot for the discrete derivative of the free-energy density,
$\dfrac{\Delta f}{\Delta \xi}$, where ${\cal N} = 1000$ interpolating systems were used for the $L_1/L_0 = 100$ run. The final precision for this
observable is comparable to that achieved for $\braket{T_{01}^{[1]}}_{\xi=1}$. For both quantities, spatial 
finite-volume effects remain negligible within the error bars for $g_0^2 \ge 0.35$, but they grow to approximately $1.8\%$ in the deep weak-coupling
regime at $g_0^2 = 0.1$. 

\subsection{$\braket{T_{01}^{[1]}}_{\xi=2}$ and $\braket{T_{00}^{\{2\}}-T_{11}^{[3]}}_{\xi=2}$}
\begin{figure}[htb]
    \center
    \includegraphics[width=0.8\linewidth]{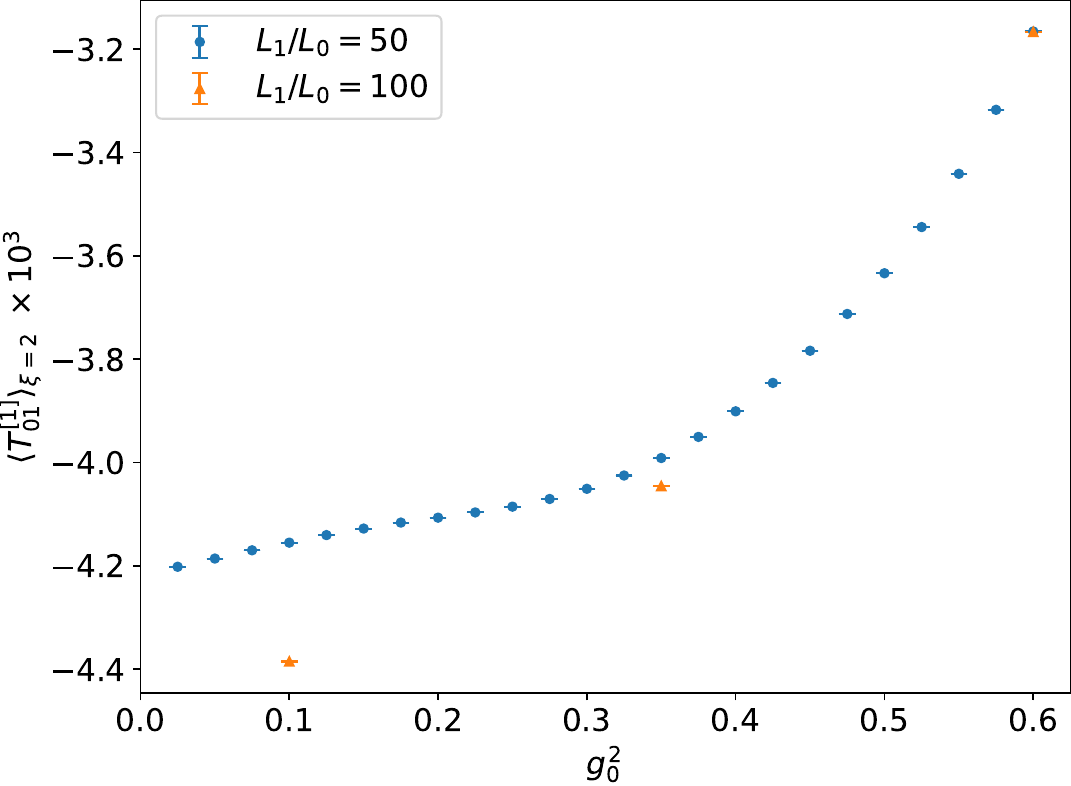}\\
    \includegraphics[width=0.8\linewidth]{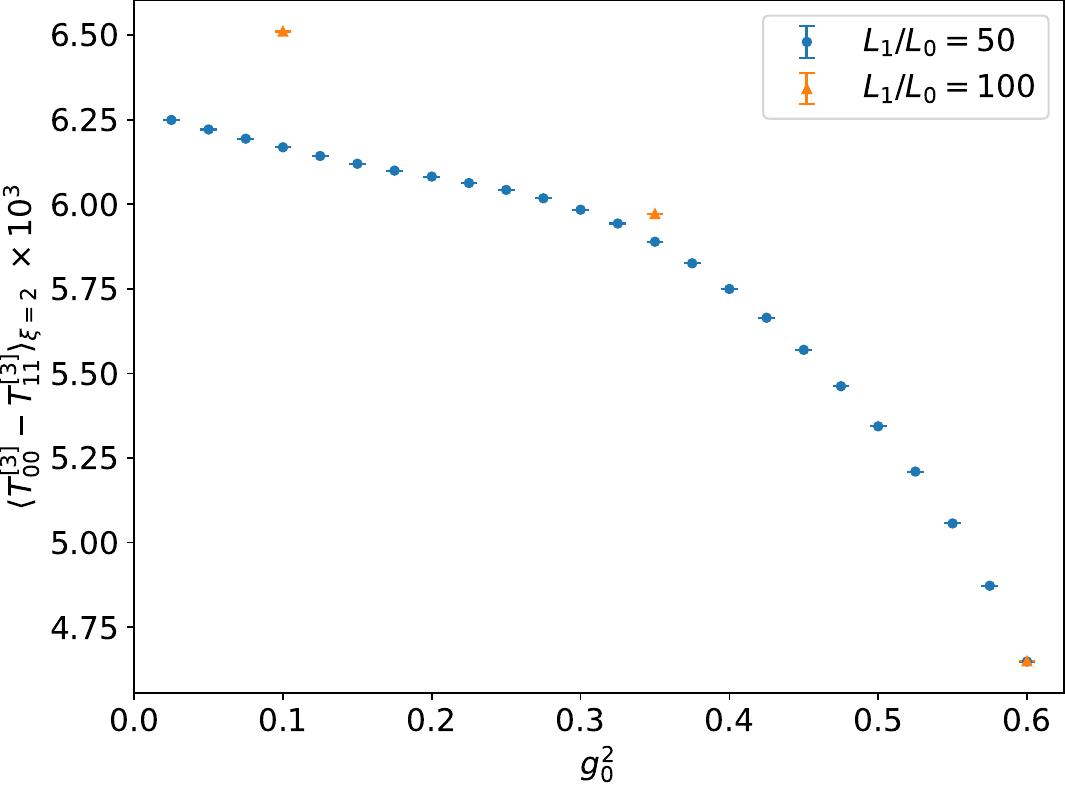}
    \caption{Comparison of the results at $L_1/L_0 = 50$ (blue-circle points) and at 100 (orange-triangle points) for
    $\braket{T_{01}^{[1]}}_{\xi=2}$ (top) and $\braket{T_{00}^{[3]}-T_{11}^{[3]}}_{\xi=2}$ (bottom) as a function of $g_0^2$ for $L_0/a=6$.}
    \label{fig:finV_T01_T00_T11}
\end{figure}

The expectation value of the off-diagonal component of the energy-momentum tensor, $\braket{T_{01}^{[1]}}_{\xi=2}$, is displayed in the
top panel of Fig.~\ref{fig:finV_T01_T00_T11}. At an aspect ratio of $L_1/L_0 = 50$, the relative precision of these measurements reaches
approximately $0.1\permil$, based on a statistics of a few $10^9$ Monte Carlo sweeps per data point. For the larger aspect
ratio of $L_1/L_0 = 100$, the precision is of the order of $0.3\permil$, obtained from a statistical sample of a few $10^8$
sweeps for each coupling value. 

An analogous plot is presented in the bottom panel of Fig.~\ref{fig:finV_T01_T00_T11} for the traceless diagonal component,
$\braket{T_{00}^{[3]} - T_{11}^{[3]}}_{\xi=2}$. The achieved numerical precision is comparable, yielding approximately 
$0.1\permil$ at $L_1/L_0 = 50$ and $0.3\permil$ at $L_1/L_0 = 100$. For both observables, spatial finite-volume effects remain entirely
negligible within the statistical precision at $g_0^2 = 0.6$. However, as the correlation length grows, they increase to approximately $1.4\%$
at $g_0^2 = 0.35$ and become highly pronounced, reaching $5.4\%$, in the deep weak-coupling regime at $g_0^2 = 0.1$. 

\subsection{The renormalization constants $Z_{_T}$ and $z_{_T}$}
\begin{figure}[htb]
    \centering
    \includegraphics[width=0.8\linewidth]{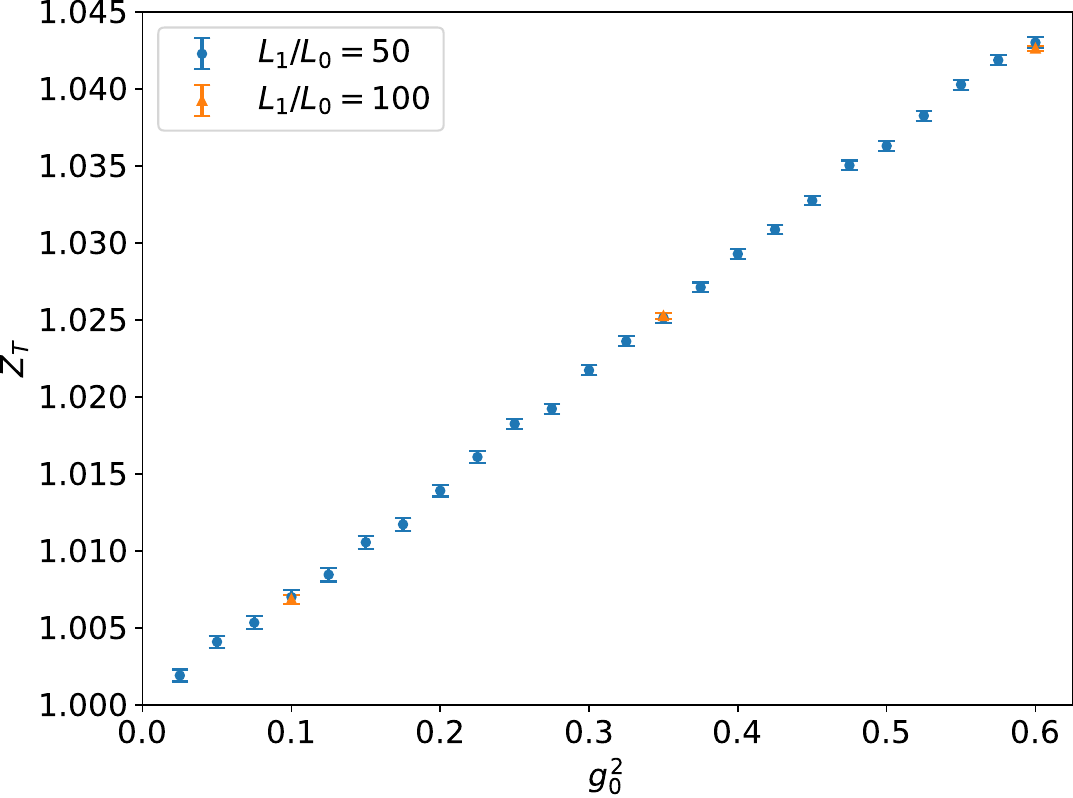}
    \\
    \includegraphics[width=0.8\linewidth]{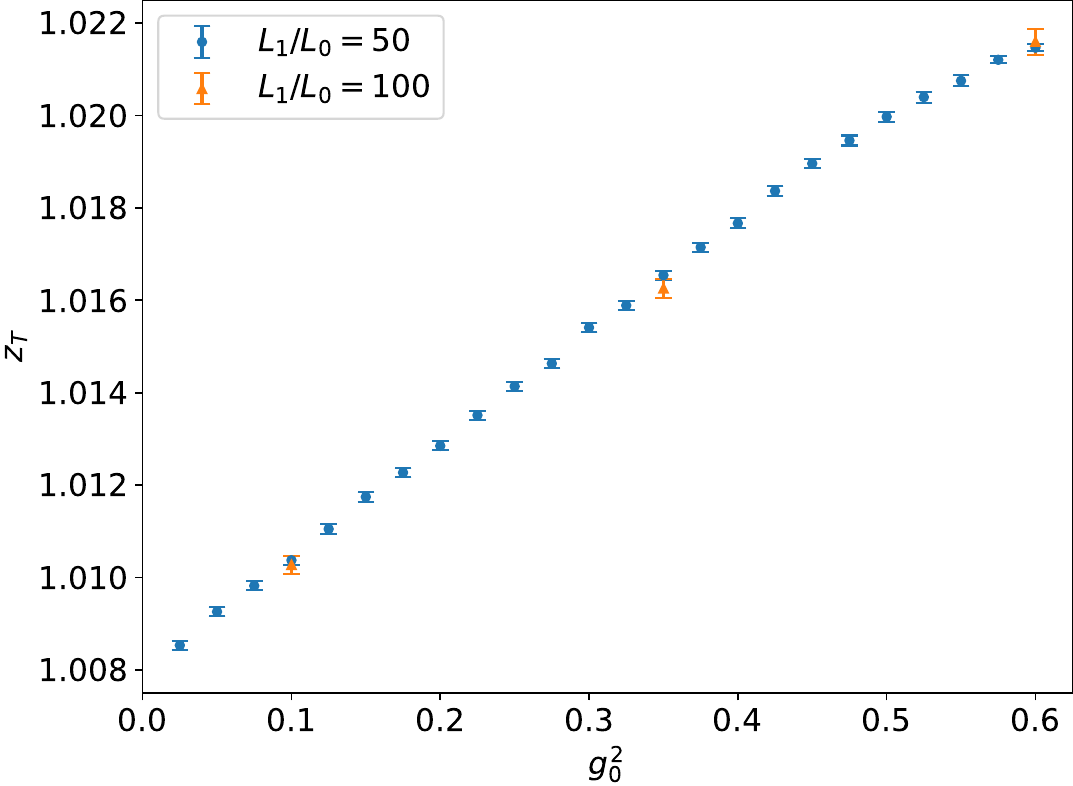}
    \caption{Comparison of the renormalization constants computed at aspect ratios $L_1/L_0 = 50$ (blue circles) and $L_1/L_0 = 100$ (orange
      triangles) as a function of the bare coupling $g_0^2$ for $L_0/a=6$. The top panel shows the overall renormalization constant $Z_T$, while the
      bottom panel shows the relative constant $z_T$.} 
    \label{fig:finV_ZT_zT}
\end{figure}

The overall renormalization constant $Z_T$ of the energy-momentum tensor is shown in the top panel of Fig.~\ref{fig:finV_ZT_zT}. The
relative numerical precision achieved for this constant is approximately $0.3$--$0.45\permil$ at $L_1/L_0 = 50$ and $0.18$--$0.3\permil$ at
$L_1/L_0 = 100$. Remarkably, because the finite-volume effects affecting $\braket{T_{01}^{[1]}}_{\xi=1}$ and the free-energy discrete
derivative $\Delta f /\Delta \xi$ at $\xi = 1$ cancel each other out, the resulting systematic finite-size effects on $Z_T$
remain entirely negligible within our statistical uncertainties across the entire range of couplings. This behavior is not unexpected, since
Ward identities holds in finite volume. 

The relative renormalization constant $z_T$ is presented in the bottom panel of Fig.~\ref{fig:finV_ZT_zT}. Its precision is roughly
$0.08$--$0.12\permil$ at $L_1/L_0 = 50$ and $0.2$--$0.28\permil$ at $L_1/L_0 = 100$. An analogous non-trivial cancellation takes place between
the finite-volume corrections of the underlying observables, $\braket{T_{01}^{[1]}}_{\xi=2}$ and
$\braket{T_{00}^{[3]} - T_{11}^{[3]}}_{\xi=2}$. This ensures that the net spatial finite-volume effects on $z_T$ are likewise suppressed and statistically
negligible.

\section{Perturbative improvement}\label{appPT}
In this Appendix, we present the one-loop perturbative computation of the traceless components of the energy-momentum tensor on the lattice for the
$O(N)$ non-linear sigma model with arbitrary $N$. These analytical results are instrumental for introducing an improved definition of the
renormalization constants, allowing a more precise non-perturbative determination. The calculation follows a standard setup, with the only
specific feature of implementing shifted boundary conditions on the spin fields. 

By taking into account the unit-norm constraint of the spin field, we integrate out the last component, $\sigma_{\!_N} (x)$, in the partition
function defined by Eqs.~(\ref{eq:Zpart}) and~(\ref{eq:latt_act}). The partition function can thus be rewritten in terms of the field
$\vec{\pi}(x)$, which is given by the first $(N-1)$ components of $\vec{\sigma}(x)$, as follows
\begin{equation}\label{Zfieldpi}
  {\cal Z} (L_0,\xi) = \int {\cal D}\vec\pi\,\prod_x \dfrac{1}{2 \sqrt{1-\vec \pi^2 (x)}} \;  e^{-S}.
\end{equation}
The product in the integrand comes from re-expressing the $\delta$-function constraint $\delta(\vec{\sigma}(x)^2 - 1)$ as
\begin{equation} 
 \! \delta\!\left[\!  \left(\! \sigma_{\!_N}\!(x)\! -\! \sqrt{1 \!-\! \vec{\pi}^2(x)}  \right) \! \!
    \left(\! \sigma_{\!_N}\!(x) \! +\! \sqrt{1\!-\! \vec{\pi}^2(x)}  \right) \! \right] ,
\end{equation}
under the choice of performing the perturbative expansion around the classical vacuum configuration $(0,\ldots,0,1)$. Accordingly, the field
$\vec{\pi}(x)$ parameterizes the fluctuations around this vacuum state. To set up perturbation theory, we rescale the field as
$\vec{\pi}(x) \rightarrow g_0 \vec{\pi}(x)$, recasting the partition function as
\begin{equation}\label{Zpert}
  {\cal Z} (L_0,\xi) = \int {\cal D}\vec\pi\, \;  e^{-S_{eff}},
\end{equation}
where the effective lattice action is given by
\begin{equation}\label{Seff}
  \begin{split}
  S_{eff}=- \sum_{x,\mu}~\bigg[&
    \vec\pi(x)\cdot \vec\pi(x+a\hat{\mu}) +  \\
   \dfrac{1}{g_0^2} \sqrt{1-g_0^2\, \vec \pi^2 (x)}& \sqrt{1-g_0^2\,\vec \pi^2 (x+a\hat{\mu})}\, \bigg] + \\
 \dfrac{1}{2} \sum_x \log &\left[ 1- g_0^2\, \vec \pi^2 (x)\right]  .
 \end{split}
\end{equation}
Considering shifted boundary conditions along the compact direction and an infinite lattice in the spatial one, the Fourier transform
is written as 
\begin{equation} 
  \vec \pi (x)\! =\!\! \int_{p_\xi}\!\!\! \tilde{\vec\pi} (p)\,  e^{i p x} \;\; \mbox{with} \;
  \int_{p_\xi}\!\!\! =\dfrac{1}{L_0}\!\sum_n \! \int_{-\frac{\pi}{a}}^{\frac{\pi}{a}}\! \dfrac{dp_1}{2 \pi} 
\end{equation}
where $p_0 = 2\pi n_0/L_0 - p_1\xi$ and $n_0 = 0, \dots, L_0/a-1$. Expanding perturbatively the action $S_{eff}$ in $g_0^2$, the free-field
propagator is given by 
\begin{equation}
    \Braket{\pi_a(x)\pi_b(y)} = \int_{p_\xi} \dfrac{\delta_{ab}}{D(p)} e^{ip(x-y)},
\end{equation}
where $\delta_{ab}$ is a Kronecker delta and $D(p) = \hat{p}_0^2 + \hat{p}_1^2$ with $\hat{p}_\mu = 2/a\sin{\left( \frac{a p_\mu}{2} \right)}$.
At one-loop order we have two interaction vertices given by
\begin{eqnarray}
  V_1 &=& \dfrac{1}{8} \int^* _{p_\xi, q_\xi, k_\xi, r_\xi}\hspace{-1.3cm}  \,
  \tilde\pi_a (p)\, \tilde\pi_a (q)\, \tilde\pi_b (k)\, \tilde\pi_b (r)\, D(p+q)  \qquad \\
  M_1 &=& -\dfrac{1}{2} \int^*_{p_\xi,q_\xi} \tilde\pi_a (p)\, \tilde\pi_a (q) \, .
\end{eqnarray}
with the $*$ standing for a $\delta$ constraint on the sum of the integration momenta.
Following the same lines as above, we have the following expressions for the lattice energy-momentum tensor at one-loop order in perturbation
theory
\begin{equation}
\begin{aligned}[t]
&T_{01}^{[1]} = \nabla^{\rm s}_0\vec{\pi}(x) \cdot  \nabla^{\rm s}_1\vec{\pi}(x) + \\
&\quad \dfrac{1}{4} g_0^2
\left[ \nabla^{\rm s}_0\vec{\pi}(x)^2 \right] \left[ \nabla^{\rm s}_1\vec{\pi}(x)^2 \right]  +O(g_0^4) \, .
\end{aligned}
\end{equation}
\begin{equation}
\begin{aligned}[t]
  & T_{00}^{[3]} - T_{11}^{[3]} =
   \left[ \nabla^{\rm s}_0\vec{\pi}(x) \right]^2 - \left[ \nabla^{\rm s}_1\vec{\pi}(x) \right]^2 + \\
 &\;\;\; \dfrac{1}{4}  g_0^2 \left[ 
  \left[ \nabla^{\rm s}_0\vec{\pi}(x)^2 \right] ^2 \! - \left[ \nabla^{\rm s}_1\vec{\pi}(x)^2 \right] ^2 \right] +O(g_0^4)\,.
\end{aligned}
\end{equation}
  
\subsection{Perturbative computation of $Z_{_T}$}
The overall renormalization constant of the energy-momentum tensor $Z_T$ is obtained by evaluating Feynman diagrams for both the space-time
components of the energy-momentum tensor and the free-energy density. Their perturbative expansion at one-loop order can be written as 
\begin{eqnarray}
    f_\xi &=& f^{(0)}_\xi + g_0^2\, f^{(1)}_\xi +O(g_0^4), \\
    \braket{ T_{01}^{[1]} }_\xi &=& T_{01}^{(0)} + g_0^2\, T_{01}^{(1)} + O(g_0^4).
\end{eqnarray}
where the tree-level and one-loop contributions are given by
\begin{eqnarray}
  f^{(0)}_\xi &=& \dfrac{(N-1)}{2} \int_{p_\xi} \log D(p) \, , \\
  f^{(1)}_\xi &=& -\dfrac{(N-1)}{8} \, [ R_\alpha^{(1)}] ^2 \, ,\\
  T_{01}^{(0)} &=& (N-1) \,S_{01}^{(1)} \, ,\\
  T_{01}^{(1)} &=& \dfrac{(N-1)}{2} \left[ R_\alpha^{(1)}\, S_{01\alpha}^{(2)} - (1 - \frac{1}{2}R_{01}^{(1)}) S_{01}^{(1)} \right]\qquad 
  \, .
\end{eqnarray}
The lattice integrals appearing in these expressions are defined in subsection~\ref{subsec:integrals}.

In Table~\ref{tab:LPT_ZT} we collect the results for the tree-level and one-loop coefficients of the perturbative expansion of $Z_{_T}$ at
finite values of $L_0/a$ for the shift parameter $\xi=1$. We also provide the corresponding values in the 
$1/L_0\rightarrow 0$ limit~\cite{Caracciolo:2000td}. We note that if these perturbative computations were carried out at a
finite spatial volume, an additional, finite-size vertex contribution would generally need to be taken into
account~\cite{Hasenfratz:1984jk}. However, we have explicitly verified that its contribution to $Z_T$ vanishes identically at 1-loop order. 
\begin{table}[ht!]
\centering
\begin{tabular}{|c||c|c|}
\cline{1-3} 
$L_0/a$ & $Z_T^{(0),~\xi=1}$ & $Z_T^{(1),~\xi=1}$ \\ \hline
6  & 0.99996447 & 0.06836022 \\
8  & 0.99998876 & 0.06832569 \\
10 & 0.99999540 & 0.06831633 \\
12 & 0.99999778 & 0.06831299 \\
14 & 0.99999880 & 0.06831156 \\
20 & 0.99999971 & 0.06831029 \\ \hline \hline 
$1/L_0 \to 0$ & 1 & 1/$\pi$ -1/4 \\ \hline
\end{tabular}
\caption{Tree-level and one-loop coefficient of the perturbative expansion of $Z_{_T}$ at finite values of $L_0/a$ for shift $\xi =1$ and in the
  $1/L_0\rightarrow 0$ limit~\cite{Caracciolo:2000td}.}
  \label{tab:LPT_ZT}
\end{table}

\subsection{Perturbative computation of $z_{_T}$}
We then computed the 1-loop perturbative expansion of the traceless diagonal component of the energy-momentum tensor to determine the
expansion coefficients of the relative renormalization constant $z_{_T}$. We write 
\begin{equation}
  \braket{ T_{00}^{[3]} - T_{11}^{[3]} }_\xi = t_{01}^{(0)} + g_0^2\, t_{01}^{(1)}+ O(g_0^4),
\end{equation}
where the tree-level and the one-loop coefficients for a generic $O(N)$ non-linear sigma model are given by
\begin{eqnarray}
  t_{01}^{(0)} &=& (N-1) \left[ S_{00}^{(1)}- S_{11}^{(1)} \right] \, ,\\
  t_{01}^{(1)} &=& (N-1) \Big[  \frac{1}{2} R_\alpha^{(1)} \left( S_{00\alpha}^{(2)}- S_{11\alpha}^{(2)} \right) + \nonumber \\
   &-& \left( {S_{00}^{(1)}}^2 - {S_{11}^{(1)}}^2 \right) \Big] .
\end{eqnarray}
The lattice integrals appearing above are defined in subsection~\ref{subsec:integrals}. 

In Table~\ref{tab:LPT_zT} we give $z_{_T}^{(0)}$ and $z_{_T}^{(1)}$ computed with $\xi=2$ at the values of $L_0/a$ considered in this study
as well as their $1/L_0\rightarrow 0$ limit~\cite{Caracciolo:2000td}. Similarly to the case of $Z_{_T}$, any potential finite-size vertex
contribution that would arise from a finite spatial volume vanishes at the one-loop level for $z_{_T}$. 
\begin{table}[ht!]
\centering
\begin{tabular}{|c||c|c|}
\cline{1-3} 
$L_0/a$ & $z_T^{(0),~\xi=2}$ & $z_T^{(1),~\xi=2}$ \\ \hline
6  & 1.00724134 & 0.02416525 \\
8  & 1.00409039 & 0.02344935 \\
10 & 1.00262292 & 0.02311946 \\
12 & 1.00182338 & 0.02294063 \\
14 & 1.00134047 & 0.02283293 \\
20 & 1.00065741 & 0.02268098 \\ \hline \hline
$1/L_0 \to 0$ & 1 & (1-3/$\pi$)/2 \\ \hline
\end{tabular}
\caption{Tree-level and one-loop coefficient of the perturbative expansion of $z_{_T}$ at finite values of $L_0/a$ for shift $\xi =2$ and in the
  $1/L_0\rightarrow 0$ limit~\cite{Caracciolo:2000td}.}
  \label{tab:LPT_zT}
\end{table}

\subsection{Integrals}
\label{subsec:integrals}
\begin{eqnarray}
    &&R^{(1)} = \int_{p_\xi} \dfrac{1}{D(p)} \\
    &&R^{(1)}_\mu = \int_{p_\xi} \dfrac{\hat{p}^2_\mu}{D(p)} \\
    &&R^{(1)}_{\mu\nu} = \int_{p_\xi} \dfrac{\hat{p}^2_\mu\hat{p}^2_\nu}{D(p)} \\
    &&S^{(1)}_{\mu\nu} = \int_{p_\xi} \dfrac{\sin(a p_\mu) \sin(a p_\nu)}{D(p)} \\
    &&S^{(2)}_{\mu\nu\alpha} = \int_{p_\xi} \dfrac{\sin(a p_\mu) \sin(a p_\nu)\hat{p}^2_\alpha}{D^2(p)} 
\end{eqnarray}

\section{Data tables}\label{appTables}
In this Appendix, we collect the Monte Carlo data at the investigated values of $g_0^2$ and $L_0/a$. 
The aspect ratio of the spatial and the temporal extents is fixed as $L_1/L_0 = 50$.
The simulations are performed using the heat-bath updating algorithm and the standard estimator for measuring observables.

\subsection{Computation of $Z_T$}
For the estimation of the renormalization constant $Z_T$, we performed simulations with $\xi=1$.
The expectation values of the space-time component of the energy-momentum tensor $\braket{T_{01}^{[1]}}_{\xi=1}$ are reported in
Tab.~\ref{tab:T01_xi1}. 
The precision is approximately $0.1-0.5\permil$ with statistics of $3.2\times 10^8$ sweeps for each set of parameters.

The derivative of free-energy with respect to the shift $\Delta f/\Delta\xi$ is estimated using the method discussed in Section~\ref{Df/Dxi}, with
${\cal N} = 500, 800, 1000, 1250, 1600,$ and 2000 respectively for $L_0/a$ from 6 to 20. 
The values of the $\Delta f/\Delta\xi$ are reported in Tab.~\ref{tab:dfdxi}.
The collected statistics at given value of the bare coupling $g_0^2$ in units of $10^8$ sweeps are $2.5,\ 4,\ 5,\ 6.25,\ 8$ and 40
for $L_0/a = 6,\ 8,\ 10,\ 12,\ 14$ and 20, respectively. The precision is approximately $0.2-0.8\permil$.

\subsection{Computation of $z_T$}
To estimate the relative renormalization constant $z_T$, we performed numerical simulations using the shift parameter $\xi=2$. The 
expectation values $\braket{T_{01}^{[1]}}_{\xi=2}$ and $\braket{T_{00}^{[3]}-T_{11}^{[3]}}_{\xi=2}$ obtained from our Monte Carlo
runs are tabulated in Table~\ref{tab:T01_xi2} and Table~\ref{tab:T00_T11_xi2}, respectively. Since both observables were measured
simultaneously on the same gauge ensembles at each point in the parameter space, they are statistically correlated; their covariances
are provided in Table~\ref{tab:cov_T01_T00_T11_xi2}.
The collected statistics at given value of the bare coupling $g_0^2$ in units of $10^{10}$ sweeps are $0.45,\ 1,\ 1.25,\ 8,\ 10$ and 15
for $L_0/a = 6,\ 8,\ 10,\ 12,\ 14$ and 20, respectively. The precision is approximately $0.07-0.2\permil$.

\begin{table*}[ht!]
\centering
\begin{tabular}{|c||c|c|c|c|c|c|}
\hline
\multirow[c]{2}{*}{$g_0^2$} & 
\multicolumn{6}{c|}{$\braket{T_{01}^{[1]}}_{\xi=1} \times 10^3$} \\ \cline{2-7}
& $L_0/a = 6$ & $L_0/a = 8$ & $L_0/a = 10$ & $L_0/a = 12$ & $L_0/a = 14$ & $L_0/a = 20$\\ \hline
 0.025 & -13.8154(28) & -7.8005(20) & -5.0015(24) & -3.4790(14) & -2.5539(12) & -1.2519(7) \\
 0.050 & -13.7674(27) & -7.7705(20) & -4.9832(25) & -3.4634(14) & -2.5465(12) & -1.2489(7) \\
 0.075 & -13.730(3) & -7.7520(23) & -4.9705(27) & -3.4560(14) & -2.5397(14) & -1.2445(7) \\
 0.100 & -13.695(3) & -7.7357(24) & -4.9565(28) & -3.4464(16) & -2.5325(13) & -1.2420(8) \\
 0.125 & -13.667(3) & -7.7130(25) & -4.9449(25) & -3.4370(17) & -2.5260(15) & -1.2383(8) \\
 0.150 & -13.6027(29) & -7.6805(23) & -4.9235(25) & -3.4209(15) & -2.5166(14) & -1.2324(7) \\
 0.175 & -13.5264(28) & -7.6326(20) & -4.8887(23) & -3.3994(14) & -2.4997(12) & -1.2244(7) \\
 0.200 & -13.4144(25) & -7.5722(20) & -4.8523(23) & -3.3723(14) & -2.4771(10) & -1.2147(6) \\
 0.225 & -13.2922(24) & -7.5021(18) & -4.8099(18) & -3.3383(12) & -2.4533(10) & -1.2019(6) \\
 0.250 & -13.1601(21) & -7.4231(16) & -4.7529(17) & -3.3020(11) & -2.4276(9) & -1.1887(5) \\
 0.275 & -13.0232(19) & -7.3421(15) & -4.7011(17) & -3.2658(9) & -2.3988(9) & -1.1741(5) \\
 0.300 & -12.8720(17) & -7.2558(14) & -4.6459(15) & -3.2254(9) & -2.3681(9) & -1.1594(5) \\
 0.325 & -12.7205(18) & -7.1663(14) & -4.5853(16) & -3.1831(9) & -2.3354(8) & -1.1421(5) \\
 0.350 & -12.5590(17) & -7.0715(12) & -4.5224(15) & -3.1381(9) & -2.3037(8) & -1.1251(4) \\
 0.375 & -12.3883(16) & -6.9704(12) & -4.4556(14) & -3.0905(8) & -2.2706(7) & -1.1072(4) \\
 0.400 & -12.2165(14) & -6.8659(11) & -4.3884(13) & -3.0420(8) & -2.2304(7) & -1.0873(4) \\
 0.425 & -12.0312(14) & -6.7576(10) & -4.3156(12) & -2.9888(8) & -2.1899(7) & -1.0656(4) \\
 0.450 & -11.8359(15) & -6.6411(10) & -4.2377(12) & -2.9312(7) & -2.1456(6) & -1.0426(4) \\
 0.475 & -11.6270(13) & -6.5145(10) & -4.1505(10) & -2.8685(7) & -2.0990(6) & -1.0165(3) \\
 0.500 & -11.4104(13) & -6.3792(10) & -4.0559(10) & -2.8010(6) & -2.0452(6) & -0.9859(3) \\
 0.525 & -11.1685(12) & -6.2287(9) & -3.9526(10) & -2.7229(6) & -1.9832(6) & -0.9509(3) \\
 0.550 & -10.9071(12) & -6.0612(8) & -3.8331(11) & -2.6302(6) & -1.9101(5) & -0.9062(3) \\
 0.575 & -10.6094(11) & -5.8641(8) & -3.6881(10) & -2.5171(6) & -1.8177(5) & -0.84614(28) \\
 0.600 & -10.2687(11) & -5.6307(7) & -3.5099(9) & -2.3735(5) & -1.6969(5) & -0.76032(25) \\
\hline
\end{tabular}

\caption{Non-perturbative results of the space-time component of the energy-momentum tensor. The simulations are performed with the shift $\xi=1$.}   
  \label{tab:T01_xi1}
\end{table*}

\begin{table*}[ht!]
\centering
\begin{tabular}{|c||c|c|c|c|c|c|}
\hline
\multirow[c]{2}{*}{$g_0^2$} & 
\multicolumn{6}{c|}{$\frac{\Delta f}{\Delta \xi}\times 10^3$} \\ \cline{2-7}
& $L_0/a = 6$ & $L_0/a = 8$ & $L_0/a = 10$ & $L_0/a = 12$ & $L_0/a = 14$ & $L_0/a = 20$ \\ \hline
 0.025 & 13.842(5) & 7.810(3) & 5.0142(22) & 3.4830(17) & 2.5582(14) & 1.2550(5) \\
 0.050 & 13.824(5) & 7.798(3) & 5.0019(22) & 3.4773(17) & 2.5534(14) & 1.2527(5) \\
 0.075 & 13.803(5) & 7.790(4) & 4.9936(23) & 3.4727(17) & 2.5519(14) & 1.2514(5) \\
 0.100 & 13.791(5) & 7.791(4) & 4.9933(24) & 3.4705(18) & 2.5488(14) & 1.2504(5) \\
 0.125 & 13.782(5) & 7.784(4) & 4.9865(23) & 3.4675(18) & 2.5471(14) & 1.2503(5) \\
 0.150 & 13.746(5) & 7.757(3) & 4.9743(23) & 3.4585(18) & 2.5392(14) & 1.2457(5) \\
 0.175 & 13.685(5) & 7.725(3) & 4.9473(22) & 3.4394(17) & 2.5292(14) & 1.2397(5) \\
 0.200 & 13.601(4) & 7.675(3) & 4.9212(21) & 3.4205(17) & 2.5114(13) & 1.2310(5) \\
 0.225 & 13.506(4) & 7.621(3) & 4.8838(20) & 3.3889(16) & 2.4934(13) & 1.2209(5) \\
 0.250 & 13.400(4) & 7.5521(29) & 4.8389(20) & 3.3627(16) & 2.4697(13) & 1.2097(5) \\
 0.275 & 13.274(4) & 7.4827(28) & 4.7946(19) & 3.3299(16) & 2.4435(13) & 1.1973(5) \\
 0.300 & 13.152(4) & 7.4156(28) & 4.7467(19) & 3.2939(15) & 2.4212(12) & 1.1843(5) \\
 0.325 & 13.021(4) & 7.3344(27) & 4.6927(19) & 3.2582(15) & 2.3916(12) & 1.1697(5) \\
 0.350 & 12.875(4) & 7.2548(26) & 4.6377(19) & 3.2170(15) & 2.3614(12) & 1.1537(5) \\
 0.375 & 12.724(4) & 7.1628(25) & 4.5809(19) & 3.1781(15) & 2.3307(12) & 1.1374(5) \\
 0.400 & 12.574(4) & 7.0668(25) & 4.5198(18) & 3.1328(15) & 2.2954(12) & 1.1192(5) \\
 0.425 & 12.402(3) & 6.9679(24) & 4.4497(18) & 3.0822(15) & 2.2580(12) & 1.0988(4) \\
 0.450 & 12.224(3) & 6.8588(24) & 4.3769(18) & 3.0275(15) & 2.2155(12) & 1.0764(4) \\
 0.475 & 12.034(3) & 6.7486(24) & 4.2982(18) & 2.9696(15) & 2.1728(12) & 1.0520(4) \\
 0.500 & 11.825(3) & 6.6129(23) & 4.2083(18) & 2.9023(14) & 2.1225(12) & 1.0232(4) \\
 0.525 & 11.596(3) & 6.4654(23) & 4.1051(18) & 2.8278(15) & 2.0614(12) & 0.9879(4) \\
 0.550 & 11.346(3) & 6.3055(22) & 3.9871(18) & 2.7352(15) & 1.9865(12) & 0.9436(5) \\
 0.575 & 11.054(3) & 6.1082(22) & 3.8424(18) & 2.6242(15) & 1.8928(12) & 0.8812(5) \\
 0.600 & 10.710(3) & 5.8744(22) & 3.6631(18) & 2.4761(15) & 1.7695(12) & 0.7935(5) \\
\hline
\end{tabular}

\caption{Non-perturbative results of the derivative of the free-energy density with respect to the shift at $\xi=1$.}    
  \label{tab:dfdxi}
\end{table*}

\begin{table*}[ht!]
\centering
\begin{tabular}{|c||c|c|c|c|c|c|}
\hline
\multirow[c]{2}{*}{$g_0^2$} & 
\multicolumn{6}{c|}{$\braket{T_{01}^{[1]}}_{\xi=2} \times 10^3$} \\ \cline{2-7}
& $L_0/a = 6$ & $L_0/a = 8$ & $L_0/a = 10$ & $L_0/a = 12$ & $L_0/a = 14$ & $L_0/a = 20$\\ \hline
 0.025 & -4.2017(6) & -2.3614(3) & -1.51096(22) & -1.04903(8) & -0.77060(6) & -0.37747(3) \\
 0.050 & -4.1858(6) & -2.3527(3) & -1.50548(22) & -1.04505(8) & -0.76774(6) & -0.37609(3) \\
 0.075 & -4.1698(6) & -2.3440(3) & -1.49967(22) & -1.04122(8) & -0.76492(6) & -0.37469(3) \\
 0.100 & -4.1549(6) & -2.3358(3) & -1.49431(22) & -1.03737(8) & -0.76202(6) & -0.37328(3) \\
 0.125 & -4.1404(6) & -2.3272(3) & -1.48915(22) & -1.03375(8) & -0.75938(6) & -0.37192(3) \\
 0.150 & -4.1278(6) & -2.3199(3) & -1.48399(23) & -1.03040(8) & -0.75689(6) & -0.37070(4) \\
 0.175 & -4.1162(7) & -2.3132(3) & -1.47970(24) & -1.02741(8) & -0.75468(6) & -0.36961(4) \\
 0.200 & -4.1066(7) & -2.3077(3) & -1.47621(24) & -1.02482(8) & -0.75280(6) & -0.36872(4) \\
 0.225 & -4.0965(7) & -2.3021(3) & -1.47273(25) & -1.02241(8) & -0.75090(6) & -0.36774(4) \\
 0.250 & -4.0853(7) & -2.2958(3) & -1.46840(25) & -1.01925(9) & -0.74867(6) & -0.36657(4) \\
 0.275 & -4.0705(7) & -2.2871(3) & -1.46255(25) & -1.01510(8) & -0.74550(6) & -0.36490(4) \\
 0.300 & -4.0508(7) & -2.2751(3) & -1.45456(24) & -1.00924(8) & -0.74106(6) & -0.36253(4) \\
 0.325 & -4.0250(6) & -2.2590(3) & -1.44365(23) & -1.00144(8) & -0.73497(6) & -0.35926(3) \\
 0.350 & -3.9912(6) & -2.2384(3) & -1.42955(22) & -0.99120(7) & -0.72722(6) & -0.35509(3) \\
 0.375 & -3.9503(6) & -2.21347(29) & -1.41278(21) & -0.97890(7) & -0.71769(5) & -0.34997(3) \\
 0.400 & -3.9008(5) & -2.18401(27) & -1.39250(20) & -0.96419(7) & -0.70650(5) & -0.343949(29) \\
 0.425 & -3.8458(5) & -2.14978(26) & -1.36967(19) & -0.94725(6) & -0.69353(5) & -0.336878(28) \\
 0.450 & -3.7836(5) & -2.11144(25) & -1.34321(18) & -0.92790(6) & -0.67861(5) & -0.328744(27) \\
 0.475 & -3.7124(5) & -2.06810(23) & -1.31331(17) & -0.90589(6) & -0.66155(4) & -0.319234(25) \\
 0.500 & -3.6336(5) & -2.01888(22) & -1.27877(16) & -0.88028(6) & -0.64152(4) & -0.307788(25) \\
 0.525 & -3.5443(4) & -1.96221(21) & -1.23875(16) & -0.84985(5) & -0.61733(4) & -0.293229(24) \\
 0.550 & -3.4412(4) & -1.89438(21) & -1.18935(15) & -0.81146(5) & -0.58604(4) & -0.273045(23) \\
 0.575 & -3.31735(24) & -1.81028(11) & -1.12586(10) & -0.76015(5) & -0.54277(3) & -0.242834(16) \\
 0.600 & -3.16574(23) & -1.70139(9) & -1.03992(10) & -0.68823(5) & -0.48045(3) & -0.197555(15) \\
\hline
\end{tabular}

\caption{Non-perturbative results of the space-time component of the energy-momentum tensor. The simulations are performed with the shift $\xi=2$.}
  \label{tab:T01_xi2}
\end{table*}

\begin{table*}[ht!]
\centering
\begin{tabular}{|c||c|c|c|c|c|c|}
\hline
\multirow[c]{2}{*}{$g_0^2$} & 
\multicolumn{6}{c|}{$\braket{T_{00}^{[3]} - T_{11}^{[3]}}_{\xi=2}  \times 10^3$} \\ \cline{2-7}
& $L_0/a = 6$ & $L_0/a = 8$ & $L_0/a = 10$ & $L_0/a = 12$ & $L_0/a = 14$ & $L_0/a = 20$\\ \hline
 0.025 & 6.2492(9) & 3.5242(5) & 2.2586(4) & 1.56950(12) & 1.15355(9) & 0.56546(5) \\
 0.050 & 6.2211(9) & 3.5090(5) & 2.2489(3) & 1.56262(12) & 1.14850(9) & 0.56303(5) \\
 0.075 & 6.1939(9) & 3.4942(5) & 2.2391(3) & 1.55602(12) & 1.14377(9) & 0.56062(5) \\
 0.100 & 6.1684(9) & 3.4794(5) & 2.2297(3) & 1.54929(12) & 1.13869(9) & 0.55821(5) \\
 0.125 & 6.1427(9) & 3.4646(5) & 2.2209(3) & 1.54315(12) & 1.13416(9) & 0.55591(5) \\
 0.150 & 6.1198(10) & 3.4516(5) & 2.2117(4) & 1.53714(12) & 1.12985(9) & 0.55377(6) \\
 0.175 & 6.0995(10) & 3.4400(5) & 2.2044(4) & 1.53186(12) & 1.12588(10) & 0.55186(6) \\
 0.200 & 6.0818(10) & 3.4295(5) & 2.1978(4) & 1.52700(13) & 1.12231(10) & 0.55014(6) \\
 0.225 & 6.0628(10) & 3.4194(5) & 2.1912(4) & 1.52270(13) & 1.11895(10) & 0.54836(6) \\
 0.250 & 6.0425(10) & 3.4082(5) & 2.1837(4) & 1.51722(13) & 1.11495(10) & 0.54636(6) \\
 0.275 & 6.0177(10) & 3.3931(5) & 2.1738(4) & 1.51008(13) & 1.10959(10) & 0.54358(6) \\
 0.300 & 5.9840(10) & 3.3736(5) & 2.1607(4) & 1.50055(13) & 1.10246(10) & 0.53968(6) \\
 0.325 & 5.9432(10) & 3.3477(5) & 2.1435(4) & 1.48797(12) & 1.09289(9) & 0.53459(5) \\
 0.350 & 5.8894(9) & 3.3155(5) & 2.1211(3) & 1.47206(12) & 1.08071(8) & 0.52810(5) \\
 0.375 & 5.8255(9) & 3.2767(4) & 2.0950(3) & 1.45309(11) & 1.06604(8) & 0.52022(5) \\
 0.400 & 5.7497(8) & 3.2317(4) & 2.0642(3) & 1.43049(10) & 1.04891(8) & 0.51105(5) \\
 0.425 & 5.6647(8) & 3.1792(4) & 2.0291(3) & 1.40483(10) & 1.02910(8) & 0.50030(5) \\
 0.450 & 5.5698(8) & 3.1207(4) & 1.98932(28) & 1.37546(10) & 1.00658(7) & 0.48806(4) \\
 0.475 & 5.4623(7) & 3.0554(4) & 1.94402(27) & 1.34220(9) & 0.98084(7) & 0.47371(4) \\
 0.500 & 5.3437(7) & 2.9814(4) & 1.89188(26) & 1.30393(9) & 0.95088(7) & 0.45656(4) \\
 0.525 & 5.2101(7) & 2.8962(3) & 1.83219(25) & 1.25830(9) & 0.91470(7) & 0.43483(4) \\
 0.550 & 5.0568(6) & 2.7953(3) & 1.75865(24) & 1.20127(8) & 0.86811(7) & 0.40484(4) \\
 0.575 & 4.8727(4) & 2.67073(17) & 1.66447(16) & 1.12508(8) & 0.80393(5) & 0.360033(27) \\
 0.600 & 4.6488(4) & 2.50958(15) & 1.53729(16) & 1.01855(8) & 0.71154(5) & 0.292888(26) \\
\hline
\end{tabular}

\caption{Non-perturbative results of $\braket{T_{00}^{[3]}-T_{11}^{[3]}}$. The simulations are performed with the shift $\xi=2$.}   
  \label{tab:T00_T11_xi2}
\end{table*}

\begin{table*}[ht!]
\centering
\begin{tabular}{|c||c|c|c|c|c|c|}
\hline
\multirow[c]{2}{*}{$g_0^2$} & 
\multicolumn{6}{c|}{${\rm cov}\left(\braket{T_{01}^{[1]}},\ \braket{T_{00}^{[3]} - T_{11}^{[3]}}  \right)\times 10^{12}$} \\ \cline{2-7}
& $L_0/a = 6$ & $L_0/a = 8$ & $L_0/a = 10$ & $L_0/a = 12$ & $L_0/a = 14$ & $L_0/a = 20$ \\ \hline
 0.0250 & -0.4071 & -0.1006 & -0.0562 & -0.0061 & -0.0035 & -0.0011 \\
 0.0500 & -0.4421 & -0.1030 & -0.0549 & -0.0062 & -0.0036 & -0.0011 \\
 0.0750 & -0.4253 & -0.1017 & -0.0546 & -0.0062 & -0.0036 & -0.0012 \\
 0.1000 & -0.4521 & -0.1034 & -0.0555 & -0.0063 & -0.0036 & -0.0012 \\
 0.1250 & -0.4480 & -0.1070 & -0.0547 & -0.0064 & -0.0037 & -0.0012 \\
 0.1500 & -0.4818 & -0.1132 & -0.0608 & -0.0068 & -0.0039 & -0.0013 \\
 0.1750 & -0.5251 & -0.1228 & -0.0663 & -0.0073 & -0.0043 & -0.0014 \\
 0.2000 & -0.5718 & -0.1361 & -0.0691 & -0.0080 & -0.0046 & -0.0015 \\
 0.2250 & -0.5501 & -0.1420 & -0.0737 & -0.0084 & -0.0048 & -0.0016 \\
 0.2500 & -0.5686 & -0.1397 & -0.0740 & -0.0084 & -0.0049 & -0.0016 \\
 0.2750 & -0.5610 & -0.1416 & -0.0742 & -0.0082 & -0.0047 & -0.0015 \\
 0.3000 & -0.5214 & -0.1336 & -0.0667 & -0.0076 & -0.0043 & -0.0014 \\
 0.3250 & -0.4836 & -0.1196 & -0.0613 & -0.0067 & -0.0039 & -0.0012 \\
 0.3500 & -0.4320 & -0.1042 & -0.0533 & -0.0059 & -0.0034 & -0.0010 \\
 0.3750 & -0.3656 & -0.0918 & -0.0460 & -0.0051 & -0.0029 & -0.0009 \\
 0.4000 & -0.3285 & -0.0769 & -0.0396 & -0.0044 & -0.0025 & -0.0007 \\
 0.4250 & -0.2756 & -0.0678 & -0.0355 & -0.0037 & -0.0021 & -0.0006 \\
 0.4500 & -0.2650 & -0.0592 & -0.0288 & -0.0032 & -0.0018 & -0.0005 \\
 0.4750 & -0.2111 & -0.0496 & -0.0250 & -0.0028 & -0.0015 & -0.0005 \\
 0.5000 & -0.1873 & -0.0424 & -0.0214 & -0.0024 & -0.0013 & -0.0004 \\
 0.5250 & -0.1621 & -0.0366 & -0.0184 & -0.0020 & -0.0011 & -0.0003 \\
 0.5500 & -0.1310 & -0.0325 & -0.0154 & -0.0017 & -0.0009 & -0.0003 \\
 0.5750 & -0.0435 & -0.0074 & -0.0058 & -0.0014 & -0.0005 & -0.0001 \\
 0.6000 & -0.0351 & -0.0050 & -0.0047 & -0.0010 & -0.0004 & -0.0001 \\
\hline
\end{tabular}

\caption{The covariance between $\braket{T_{01}^{[1]}}$ (Tab.~\ref{tab:T01_xi2}) and $\braket{T_{00}^{[3]}-T_{11}^{[3]}}$
  (Tab.~\ref{tab:T00_T11_xi2}) at $\xi=2$.}    
  \label{tab:cov_T01_T00_T11_xi2}
\end{table*}

\clearpage

\bibliographystyle{JHEP}
\bibliography{ref}
\end{document}